\documentclass[11pt,a4paper]{article}
\usepackage[T1]{fontenc}
\usepackage{lmodern}
\usepackage[a4paper,margin=2.35cm]{geometry}
\usepackage{microtype}
\usepackage{amsmath, amsfonts, amssymb}
\usepackage{tabularx, graphicx, xcolor}
\usepackage{siunitx}
\usepackage{enumitem}
\usepackage{lineno}
\usepackage{svg}
\usepackage{ragged2e}
\usepackage{authblk}
\usepackage{xurl}
\usepackage[
    colorlinks=true,
    linkcolor=black,
    citecolor=blue!55!black,
    urlcolor=blue!55!black,
    pdfauthor={Alejandro S. Gomez, Rafael A. Molina, Pablo Burset, and Yuriko Baba},
    pdftitle={Multiple chiral Majorana states in proximitized magnetic topological insulator heterostructures}
]{hyperref}
\usepackage{orcidlink}
\renewcommand{\vec}[1] {\mathbf{#1}}

\usepackage{mathtools}
\newcommand{\bra}[1] {\langle #1{\rvert}}
\newcommand{\ket}[1] {\lvert #1{\rangle}}

\usepackage[capitalise]{cleveref}

\usepackage[acronym,nomain,nonumberlist]{glossaries}
\glsdisablehyper
\newacronym{bz}{BZ}{Brillouin Zone}
\newacronym{bdg}{BdG}{Bogoliubov-de Gennes}
\newacronym{ti}{TI}{Topological Insulator}
\newacronym{phs}{PHS}{Particle Hole Symmetry}
\newacronym{cs}{CS}{Chiral Symmetry}
\newacronym{is}{IS}{Inversion Symmetry}
\newacronym{trim}{TRIM}{Time Reversal Invariant Momenta}
\newacronym{bcs}{BCS}{Bardeen–Cooper–Schrieffer}
\newacronym{obc}{OBC}{Open Boundary Conditions}
\newacronym{pbc}{PBC}{Periodic Boundary Conditions}
\newacronym{bhz}{BHZ}{Bernevig-Hughes-Zhang}
\newacronym{soc}{SOC}{spin-orbit coupling}
\newacronym{ql}{QL}{Quintuple Layer}
\newacronym{mzm}{MZM}{Majorana Zero Mode}
\newacronym{qah}{QAH}{Quantum Anomalous Hall}
\newacronym{qsh}{QSH}{Quantum Spin Hall}
\newacronym{2d}{2D}{two dimensional}
\newacronym{3d}{3D}{three dimensional}

\newcommand{\hGamma}[1]{\hat{\Gamma}_{#1}}

\newcommand{\llbrace}{\left\lbrace}
\newcommand{\rrbrace}{\right\rbrace}

\usepackage{empheq}
\usepackage{physics}
\usepackage[mathscr]{euscript}
\usepackage{mathrsfs}
\usepackage[thinc]{esdiff}
\usepackage{cancel}
\numberwithin{equation}{section}

\newcommand{\iem}{Instituto de Estructura de la Materia IEM-CSIC, Serrano 123, E-28006 Madrid, Spain}
\newcommand{\uam}{Department of Theoretical Condensed Matter Physics, Universidad Aut\'onoma de Madrid, 28049 Madrid, Spain}
\newcommand{\ifimac}{Condensed Matter Physics Center (IFIMAC), Universidad Aut\'onoma de Madrid, 28049 Madrid, Spain}
\newcommand{\inc}{Instituto Nicol\'as Cabrera, Universidad Aut\'onoma de Madrid, 28049 Madrid, Spain}

\begin{document}

\justifying
\pagestyle{plain}

\title{Multiple chiral Majorana states in proximitized magnetic topological insulator heterostructures}

\author[1,2]{Alejandro S. Gómez~\orcidlink{0009-0001-1117-4490}}
\author[3]{Rafael A. Molina~\orcidlink{0000-0001-5728-0734}}
\author[1,2,4]{Pablo Burset~\orcidlink{0000-0001-5726-0485}}
\author[3]{Yuriko Baba~\orcidlink{0000-0003-0647-3477}\thanks{Corresponding author: \href{mailto:yuriko.baba@csic.es}{yuriko.baba@csic.es}}}

\affil[1]{\uam}
\affil[2]{\ifimac}
\affil[3]{\iem}
\affil[4]{\inc}

\date{}
\maketitle

\begin{abstract}
Achieving robust topological superconductivity with multiple Majorana channels is a key step for scalable topological quantum computing.
To this end, we investigate magnetically doped three-dimensional topological insulator heterostructures proximitized by an $s$-wave superconductor within a fully three-dimensional extended Bernevig-Hughes-Zhang framework that explicitly accounts for vertical confinement.
We show that magnetic exchange coupling, orbital mixing, and bulk band inversion cooperate to generate effective equal-spin $p$-wave pairing channels supporting multiple chiral Majorana modes.
The number of Majorana channels is determined by the confined modes in the vertical direction and the high-Chern-number phases of the normal state, which allows us to derive an analytical criterion for the emergence of the multiple-Majorana topological superconducting phases.
The chiral Majorana modes remain robust against strong disorder and moderate symmetry-breaking perturbations.
We also demonstrate that a superconducting phase difference defining a vertical Josephson junction  acts as a tunable parameter that controls the hybridization, minigap, and effective multiplicity of the low energy Majorana channels.
Our results establish magnetic three-dimensional topological-insulator heterostructures as a promising platform for engineering multiple chiral Majorana modes.
\end{abstract}

\section{Introduction}

Topological phases of matter have become one of the central themes of condensed matter physics due to their unconventional boundary excitations and their robustness against local perturbations. Among these systems, \glspl{ti} occupy a prominent role as materials characterized by an insulating bulk and metallic boundary states protected by topology and symmetry~\cite{Hasan2010,Qi2011}. When time-reversal symmetry is broken, for instance through magnetic doping, \glspl{ti} can host the \gls{qah} effect where dissipationless chiral edge channels appear without the need for external magnetic fields~\cite{Chang2013, Yuan2023, Zhuo2024}. The combination of such topological states with superconductivity has attracted significant attention~\cite{Kayyalha2020, Uday2024,Yi2024, Balakrishnan2025} because it provides a natural route toward realizing topological superconductors and Majorana quasiparticle.

Majorana states are exotic quasiparticles that are their own antiparticles and obey non-Abelian exchange statistics~\cite{Kitaev2001,Nayak2008, Alicea2012,Beenakker2013}. Their potential application to fault-tolerant quantum computation has motivated extensive theoretical and experimental efforts to engineer platforms supporting them~\cite{Sarma2015}. Still, the unambiguous identification of topologically protected Majorana modes remains challenging after more than a decade of research~\cite{Kouwenhoven2025}. One of the earliest proposals for realizing chiral Majorana modes considered a single-channel \gls{qah} system near the topological transition and proximity-coupled to a conventional superconductor~\cite{Qi2010}. In this setup, the chiral electronic edge mode of the \gls{qah} phase fractionalizes into a chiral Majorana channel propagating along the system boundary. Since then, several works have generalized this idea to more realistic heterostructures involving conventional \(s\)-wave superconductors in proximity to \gls{qah} or quantum spin Hall systems~\cite{Fu2008, Lutchyn2010, Oreg2010, He2014, DiMiceli2023}.

Magnetically doped topological insulator heterostructures constitute a particularly promising platform for realizing topological superconductivity. In thin-film geometries, magnetic exchange fields combined with the superconducting proximity effect can generate topological superconducting phases supporting chiral or helical Majorana modes~\cite{Qi2010, Wang2015}. Furthermore, the effect of disorder~\cite{Lian2018, Huang2018, Zhang2020} and generalizations of the original single-edge proposal have been investigated in systems where several surface or subband states coexist~\cite{Wang2018, Legendre2024, Zsurka2025}. However, most previous studies rely on effective two-dimensional descriptions in which the vertical direction is integrated out and the relevant physics is encoded through projected surface states associated with the top and bottom surfaces of the magnetic \gls{ti}.

\begin{figure}
    \centering
    \includegraphics[width=\linewidth]{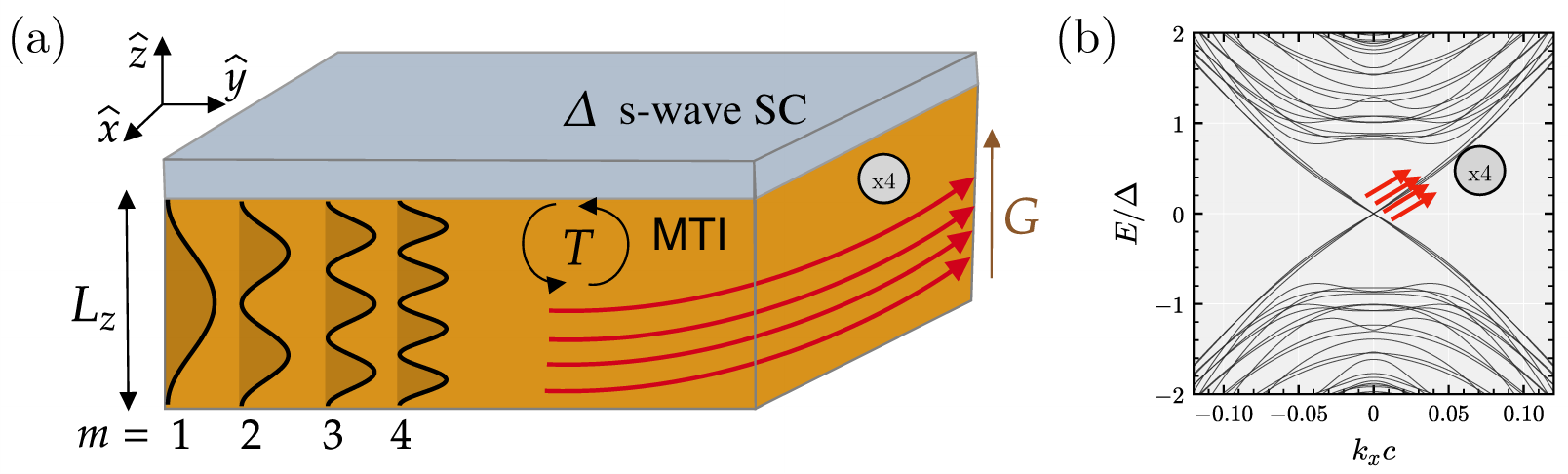}
    \caption{\textbf{Quantum-well origin of multiple chiral Majorana channels.}
    \textbf{(a)} Schematic of a magnetically doped \gls{ti} slab proximitized by an \(s\)-wave superconductor. Confinement along \(\hat z\) produces transverse modes \(m=1,\ldots,4\); the red arrows represent the resulting co-propagating chiral Majorana channels on a lateral boundary.
    \textbf{(b)} Low-energy \gls{bdg} spectrum \(E/\Delta\) versus in-plane momentum \(k_xc\) for a slab periodic along \(\hat x\) and finite along \(\hat y\) and \(\hat z\), with \(L_y/c=150\) and \(L_z/c=20\). The red arrows highlight four co-propagating subgap branches on one boundary; the opposite boundary hosts the counter-propagating partners.}
    \label{Fig1}
\end{figure}

In this work, we investigate topological superconductivity in magnetically doped three-dimensional \gls{ti} heterostructures within a fully three-dimensional framework based on an extended \gls{bhz} model~\cite{Bernevig2006, Liu2010}. In contrast to effective two-dimensional approaches, our description explicitly incorporates the vertical spatial structure and the quantum confinement along the $\hat{z}$ direction; see \cref{Fig1}. The system is proximity-coupled to a conventional $s$-wave superconductor and subjected to magnetic exchange coupling induced by magnetic doping.
We show that the interplay between the magnetic exchange field, the orbital mixing term, and the inverted bulk band structure produces effective equal-spin $p$-wave pairing channels that give rise to multiple chiral Majorana states.

Even in the absence of superconductivity, magnetically doped three-dimensional \glspl{ti} can support high-Chern-number phases characterized by multiple chiral surface channels~\cite{Zhao2020,Kezilebieke2020,Wang2021,Zhu2022}. The number of these channels depends strongly on the details of the vertical confinement, either through quantum-well subbands or through multilayer heterostructures formed by alternating magnetic and nonmagnetic layers. Here we demonstrate that, upon introducing superconducting pairing, these high-Chern phases evolve into topological superconducting phases hosting several chiral Majorana channels. Importantly, the number of Majorana modes is directly determined by the number of confined modes available in the vertical direction, allowing us to derive an analytical criterion for the appearance of the different topological phases.
To further characterize the resulting superconducting states, we employ the Majorana polarization as a numerical diagnostic capable of identifying the spatial localization and Majorana nature of the low-energy excitations. We also investigate the robustness of the chiral Majorana modes against Anderson disorder and against perturbations that explicitly break the protecting symmetries of the system. Finally, we consider a vertical Josephson junction geometry and study the evolution of the chiral Majorana states as a function of the superconducting phase difference.

\par Unlike previous surface-state-based approaches, the topological transitions we identify are driven by bulk subband inversions in the confined \gls{3d} electronic structure rather than by surface Dirac cones. Consequently, intrinsically \gls{3d} effects govern the Majorana phases.
These findings establish magnetically doped three-dimensional \gls{ti} heterostructures as a versatile platform for engineering multiple chiral Majorana states beyond the effective two-dimensional paradigm.

The remainder of the paper is organized as follows. In \cref{sec:ModelAndDiagnostics} we introduce the fully \gls{3d} model that describes proximitized magnetic topological-insulator heterostructures and we discuss the interplay between confinement, exchange fields, and superconducting proximity coupling. We also provide the definition of the main observables and diagnostic tools used subsequently.  \cref{sec:2Dlimit} analyzes the two-dimensional limit of the model and the appearance of the topological-superconducting phase as a function of the model parameters. In \cref{sec:Stacking} we study the confinement in the vertical direction and we demonstrate the emergence of multiple chiral Majorana edge modes associated with the high-Chern phases in the quantum well approximation. In \cref{sec:LinearZ} particular attention is devoted to the role of genuinely \gls{3d} effects, including inter-subband hybridization induced by the $k_z$ dispersion, which strongly reshape the topological phase diagram and the stability of the Majorana phases. In \cref{sec:disorder} we study the robustness of the Majorana states to random perturbations modeled as Anderson-like disorder terms. In \cref{sec:JJ} we study a vertical Josephson geometry which should be close to experimental applications. Finally, \cref{sec:conclusions} presents the conclusions and discusses possible experimental realizations and future extensions of this work.

\section{Model and diagnostics}\label{sec:ModelAndDiagnostics}
\subsection{BHZ model with proximity-induced superconductivity}
The magnetic \gls{ti} is described using the four band \gls{bhz}-type model for Bi$_2$Se$_3$ introduced in Refs.~\cite{Zhang2009, Liu2010}.
The normal-state basis is
\[
\llbrace
\ket{P1_z^+,\uparrow},
\ket{P2_z^-,\downarrow},
\ket{P1_z^+,\downarrow},
\ket{P2_z^-,\uparrow}
\rrbrace ,
\]
where \(P1_z^+\) and \(P2_z^-\) denote orbitals with opposite parity and
\(s= \uparrow,\downarrow\) labels the spin projection along \(\hat z\).
In this basis we consider Pauli matrices \(\hat s_i\) and \(\hat\sigma_i\) acting on spin and orbital degrees of freedom, respectively.

The unperturbed Hamiltonian is constrained by
time-reversal symmetry, inversion symmetry, and the three-fold rotation
symmetry around the \(\hat z\) axis; respectively represented by
\[
\mathcal T=i\hat{s}_y \hat{\sigma}_z\mathcal K,\qquad
\mathcal P=\hat{\sigma}_z,\qquad
C_3=\exp\!\left(i\frac{\pi}{3}\hat{s}_z \hat{\sigma}_z\right),
\]
where \(\mathcal K\) denotes complex conjugation. The magnetic
\gls{ti} regime is then obtained by adding two symmetry-breaking
perturbations: a Zeeman splitting \(G\), which breaks time-reversal symmetry; and an
off-diagonal inter-orbital hopping \(T\), which breaks inversion symmetry. These two
terms will play a central role once superconductivity is included since
their combined action generates an effective equal-spin chiral pairing
component in the proximitized system.

We now consider that superconductivity is proximity-induced by a conventional superconductor with singlet $s$-wave pairing. We express the Hamiltonian in a \gls{bdg} form by working in the Nambu basis
\[
\Psi_{\mathbf k}=
\bigl(
\hat{a}_{\mathbf k,\uparrow},
\hat{b}_{\mathbf k,\downarrow},
\hat{a}_{\mathbf k,\downarrow},
\hat{b}_{\mathbf k,\uparrow},
\hat{a}^\dagger_{-\mathbf k,\uparrow},
\hat{b}^\dagger_{-\mathbf k,\downarrow},
\hat{a}^\dagger_{-\mathbf k,\downarrow},
\hat{b}^\dagger_{-\mathbf k,\uparrow}
\bigr)^T ,
\]
where \(\hat{a}_{\mathbf k, s}\) and \(\hat{b}_{\mathbf k, s}\) label the \(P1_z^+\) and \(P2_z^-\) orbitals, respectively.
In this basis we define
\[
\hGamma{ijk}=\hat\tau_i\otimes\hat s_j\otimes\hat\sigma_k ,
\]
with \(\hat\tau_i\) Pauli matrices acting on Nambu (electron-hole) space.
Then, the full \gls{bdg} Hamiltonian used throughout the paper is \cite{Liu2010, Zhang2009}
\begin{equation}
\begin{aligned}
\mathcal H_{\rm BdG}(\mathbf k)
={}&
\varepsilon_{\mathbf k}\,\hGamma{z00}
+\mathcal M_{\mathbf k}\,\hGamma{z0z}
+\Delta (z)\,\hGamma{yyz}
+T\,\hGamma{zyy}
+G\,\hGamma{zzz}
\\
&+\frac{A}{a}\sin(k_xa)\,\hGamma{00x}
+\frac{A}{a}\sin(k_ya)\,\hGamma{zzy}
+\frac{B}{c}\sin(k_zc)\,\hGamma{0yy}.
\end{aligned}
\label{eq:FullBdG}
\end{equation}
Here, \(\Delta\) is the proximity-induced spin-singlet \(s\)-wave pairing,
\(G\) is the Zeeman splitting associated with magnetic order \cite{Yu2010, Wang2013}, \(T\) is an
off-diagonal inter-orbital hopping, and \(B\) controls the linear-in-\(k_z\)
orbital-mixing term of the \gls{3d} \gls{ti} model, see the setup scheme in \cref{Fig1}(a).
The inter-orbital coupling $T$ is obtained due to the breaking of the inversion symmetry in the growth direction, which is caused by the substrate on which the films are typically grown. It can be particularly important in thin slabs \cite{Shan2010, Zhang2010, Zhang2014}.
The lattice regularized scalar dispersion and Dirac mass are
\begin{align}
\varepsilon_{\mathbf k}
&=C_0+\mu+
\frac{2C_2}{a^2}
\left[ 2-\cos(k_xa)-\cos(k_ya) \right]
+
\frac{2C_1}{c^2} \left[ 1-\cos(k_zc) \right],
\label{eq:dispersion2D}
\\
\mathcal M_{\mathbf k}
&=
M_0+ \frac{2M_2}{a^2}
\left[ 2-\cos(k_xa)-\cos(k_ya) \right]
+
\frac{2M_1}{c^2} \left[ 1-\cos(k_zc) \right].
\label{eq:MassFull3D}
\end{align}
The constants \(a\) and \(c\) denote the lattice spacings in the in-plane and out-of-plane directions, respectively, such that the anisotropy of the unit cell in this family of materials is taken into account.
In our numerical calculations we use the induced pairing amplitude \(\Delta\) as the energy unit and the out-of-plane lattice spacing \(c\) as the length unit.
The parameters used to represent Bi$_2$Se$_3$ and related materials are
\(M_0/\Delta=-2.8\), \(M_1/(\Delta c^2)=25\), \(M_2/(\Delta c^2)=141.5\), \(A/(\Delta c)=20.5\), \(C_0/\Delta=-0.068\), \(C_2/(\Delta c^2)=49.0\), and \(a/c\simeq2.07\). These ratios correspond to the \textit{ab initio} Bi$_2$Se$_3$ values \(M_0=-0.28\,\mathrm{eV}\), \(M_2=56.6\,\mathrm{eV}\,\text{\AA}^2\), \(A=4.1\,\mathrm{eV}\,\text{\AA}\), \(C_0=-0.0068\,\mathrm{eV}\), \(C_2=19.6\,\mathrm{eV}\,\text{\AA}^2\), and \(a=4.14\,\text{\AA}\)~\cite{Zhang2009},
with \(\Delta c=0.2 \,\mathrm{eV} \text{~\AA} \). Reducing the length-energy scaling factor \(\Delta c\) increases the computational cost without qualitatively affecting our results;
see Appendix \cref{Fig1Ap} for a detailed discussion of the dependence of the topological phases on $\Delta c$.
The non-magnetic \gls{ti} belongs to class AII in Altland–Zirnbauer symmetry classification~\cite{Schnyder2008, Chiu2016}. Once the Zeeman exchange field is introduced, time-reversal symmetry is broken and the normal state problem falls into class A, while the superconducting \gls{bdg} Hamiltonian belongs to class D.

In the following, we consider various geometrical configurations, all of which can be obtained as limiting cases of the Hamiltonian, \cref{eq:FullBdG}.
The strictly two-dimensional limit is obtained by setting \(C_1=M_1=B=0\), while the quantum-well regime [\Cref{Fig1}(a)] is obtained by setting \(M_1\neq0\) and $C_1=B=0$, which leaves the transverse modes approximately labeled by discrete momenta \(k_z^{(m)}\).
Finally, the more realistic \gls{3d} slab is obtained by restoring \(B\neq0\), which mixes the vertical subbands and modifies the multi-channel structure.

\subsection{Diagnostics for the chiral Majorana modes}

To characterize the low-energy boundary states we use a combination of spectral, charge, localization, and Majorana polarization diagnostics.
In a \gls{bdg} system, particle-hole symmetry is represented by the antiunitary operator
\begin{equation}
    \mathcal C=\Gamma_{x00}\mathcal K ,
\end{equation}
which satisfies
\begin{equation}
    \mathcal C\mathcal H_{\rm BdG}(\mathbf k)\mathcal C^{-1}
    =
    -\mathcal H_{\rm BdG}(-\mathbf k).
\end{equation}
A strictly zero-energy state at a particle-hole-invariant momentum is a Majorana state when it is self-conjugate under \(\mathcal C\),
\begin{equation}
    \mathcal C\ket{\psi}
    =
    e^{i\theta}\ket{\psi}.
\end{equation}
For chiral boundary modes at finite \(k_x\), however, particle-hole symmetry relates states at opposite momenta, \(k_x\) and \(-k_x\). Therefore, away from the exact zero-energy crossing, self-conjugation should be understood as a local Majorana polarization diagnostic rather than as a standalone topological criterion.
Consequently, for an eigenstate labeled by $\nu$ and written in real space as
\begin{equation}
    \Psi_\nu(\mathbf r)
    =
    \begin{pmatrix}
    u_\nu(\mathbf r)\\
    v_\nu(\mathbf r)
    \end{pmatrix},
\end{equation}
where \(u_\nu\) and \(v_\nu\) are four-component electron and hole spinors, we define the local probability density
\begin{equation}
    \rho_\nu(\mathbf r)
    =
    \sum_{\alpha=1}^{4}
    \left(
    |u_{\nu,\alpha}(\mathbf r)|^2
    +
    |v_{\nu,\alpha}(\mathbf r)|^2
    \right) ,
\end{equation}
and the \textit{local Majorana polarization}~\cite{Sticlet2012, Sedlmayr2015}
\begin{equation}
    {\rm MP}_\nu(\mathbf r)
    =
    2\sum_{\alpha=1}^{4}
    u_{\nu,\alpha}(\mathbf r)
    v_{\nu,\alpha}(\mathbf r),
    \label{eq:local-majorana-polarization}
\end{equation}
where $\alpha=1,\dots,4$ labels the spinor component index.
The local Majorana polarization is a complex quantity that defines the spatial dependence of the Majorana character of a given state.
A pure Majorana state has an aligned Majorana polarization vector in the region where the state is localized, whereas a topological state might exhibit a high ${\rm MP}_\nu$ locally, but without any spatial correlation. Therefore, it is useful to define the \textit{Majorana polarization} as being restricted to a spatial region \(\Omega\)~\cite{Sedlmayr2015, Karoliya2025}, namely,
\begin{equation}
    \chi_\nu(\Omega)
    =
    \frac{
    \left|
    \sum_{\mathbf r\in\Omega}
    {\rm MP}_\nu(\mathbf r)
    \right|^2
    }{
    \left[
    \sum_{\mathbf r\in\Omega}
    \rho_\nu(\mathbf r)
    \right]^2
    }.
    \label{eq:regional-majorana-polarization}
\end{equation}
This quantity satisfies \(0\leq\chi_\nu(\Omega)\leq1\), i.e., values close to one indicate that the state is locally close to a self-conjugate Majorana mode in the region \(\Omega\), while values close to zero indicate a predominantly ordinary Andreev-like excitation. In ribbon geometries we evaluate
\cref{eq:regional-majorana-polarization} separately on the left and right edges, that is,
\begin{equation}
    \chi_\nu^{L/R}
    \equiv
    \chi_\nu(\Omega_{L/R}),
\end{equation}
where \(\Omega_L\) and \(\Omega_R\) denote narrow strips near the two opposite boundaries. As a result, $\chi_\nu^{L/R}$ identifies both the Majorana character and the edge localization of each low-energy branch.
Note that in the limiting case of the ideal quantum well basis with \(B=0\) we replace the subscript $\nu \to m$, with $m$ the index of the transverse quantum-well mode.

For fully finite \gls{3d} geometries, we further use the phase winding of the local Majorana polarization vector~\cite{Sedlmayr2015}. Writing
\begin{equation} \label{eq:localMP}
    \mathbf {MP}_\nu(\mathbf r)
    =
    \left[
    \operatorname{Re} \{ {\rm MP}_\nu(\mathbf r) \},
    \operatorname{Im} \{ {\rm MP}_\nu(\mathbf r) \}
    \right],
\end{equation}
we assign to a boundary-localized low-energy state the winding number
\begin{equation}
    \gamma_\nu
    =
    \frac{1}{2\pi}
    \oint_{\mathtt{C}}
    d\ell\,
    \partial_\ell
    \arg
    \left[
    \sum_{\mathbf r\in S(\ell)}
    {\rm MP}_\nu(\mathbf r)
    \right],
    \label{eq:majorana-polarization-winding}
\end{equation}
where $\mathtt{C}$ is a closed contour following the boundary of the sample and \(S(\ell)\) denotes a small transverse section around the contour point \(\ell\). States with \(\gamma_\nu\simeq+1\) and \(\gamma_\nu\simeq-1\) correspond to opposite windings of the local Majorana polarization vector and are used below to distinguish the chirality of \gls{3d} boundary modes in disordered samples.

As a complementary diagnostic we compute the charge expectation value,
\begin{equation}
    \langle Q_\nu \rangle
    =
    \bra{\psi_\nu}
    e\Gamma_{z00}
    \ket{\psi_\nu}.
    \label{eq:charge-expectation}
\end{equation}
A chiral Majorana branch is expected to have approximately balanced electron
and hole weights, and therefore \(|Q_\nu|\ll e\). We emphasize, however, that
a small charge expectation value alone is not sufficient to identify a
topological Majorana mode since it is also a property of a general Andreev state; in the following we use it only together with the
bulk gap, the edge localization, the spectral flow across zero energy, and the Majorana-polarization indicator.

\section{Two-dimensional limit for the proximitized magnetic TI}
\label{sec:2Dlimit}

We first isolate the strictly two-dimensional problem by setting \(C_1=M_1=B=0\) in \cref{eq:FullBdG}. This choice removes the out-of-plane dispersion and the linear \(k_z\) orbital-mixing term, leaving a class D superconducting \gls{bhz} Hamiltonian for a single layer. The two-dimensional limit then provides the elementary chiral building block that, in the following sections, will be replicated by quantum-well subbands in the slab geometry.

The key mechanism to achieve a superconducting topological phase can be seen from the anomalous Green function. Appendix~\ref{app:2d-derivation} gives the complete Clifford algebra derivation of the secular equation of the eight component \gls{bdg} matrix. The corresponding anomalous Green function contains, at low energy, the equal-spin component
\begin{equation}
\mathcal{G}_{\,e\uparrow,\,h\uparrow}(E{\simeq}0,\mathbf k)
      \;\propto\;
      i\,G\,T\,\Delta\,\bigl(k_x\pm i k_y\bigr).
\label{eq:effective-pwave}
\end{equation}
Thus, the Zeeman field \(G\), the inter-orbital hopping \(T\), and the parent
\(s\)-wave order parameter \(\Delta\) combine to generate an effective spin-polarized \(p_x\pm ip_y\) pairing channel that is responsible for the chiral superconducting phase.

Since a change in the topological phase requires the gap to close, we determine the zeros of the gap $\delta E_{\rm min}$ to identify the phase map of the model in the parameter space.
The corresponding bulk phase boundaries follow by setting \(E=0\) in the secular equation of the two-dimensional \gls{bdg} Hamiltonian, which leads to the conditions
\begin{align}\label{eq:GapClosings}
\text{(i)}\;&\; \sin^2(k_xa)+\sin^2(k_ya)=0
          \;\Longleftrightarrow\;
          \mathbf{k}=\Lambda_i \;(\text{TRIM}),\nonumber\\
\text{(ii)}&\;\bigl[{\varepsilon}_{\Lambda_i}^{2}+\Delta^{2} -\mathcal{M}_{\Lambda_i}^{2}-G^{2}-T^{2}\bigr]^{2}
=  4\mathcal{M}_{\Lambda_i}^{2}\!\bigl(G^{2}-\Delta^{2}\bigr) +4G^{2}T^{2}.
\end{align}
Hence, the gap closes only at the \gls{trim} $\Lambda_i$ and for the parameters that fulfill condition (ii).
Near the \(\Gamma\) point we write
\({\varepsilon}_{\Gamma}=C_{0}+\mu\equiv\tilde{\mu}\), so that \cref{eq:GapClosings} gives analytical boundaries in the \(\{G,\Delta,T,\tilde{\mu}\}\) parameter space. In particular, for the Zeeman splitting $G$, which is tunable via the magnetic doping and external fields, the phase boundaries are set by the solutions of
\begin{equation}\label{eq:boundary-C0fixed-2}
G^{2}=\Delta^{2}+\tilde{\mu}^{2}+\mathcal{M}_{\Gamma}^{2}+T^{2}
      \;\pm\;2\sqrt{\tilde{\mu}^{2}\!\bigl(\mathcal{M}_{\Gamma}^{2}+T^{2}\bigr)+T^{2}\Delta^{2}}~.
\end{equation}

\begin{figure}[t]
    \centering
    \includegraphics[width=\linewidth]{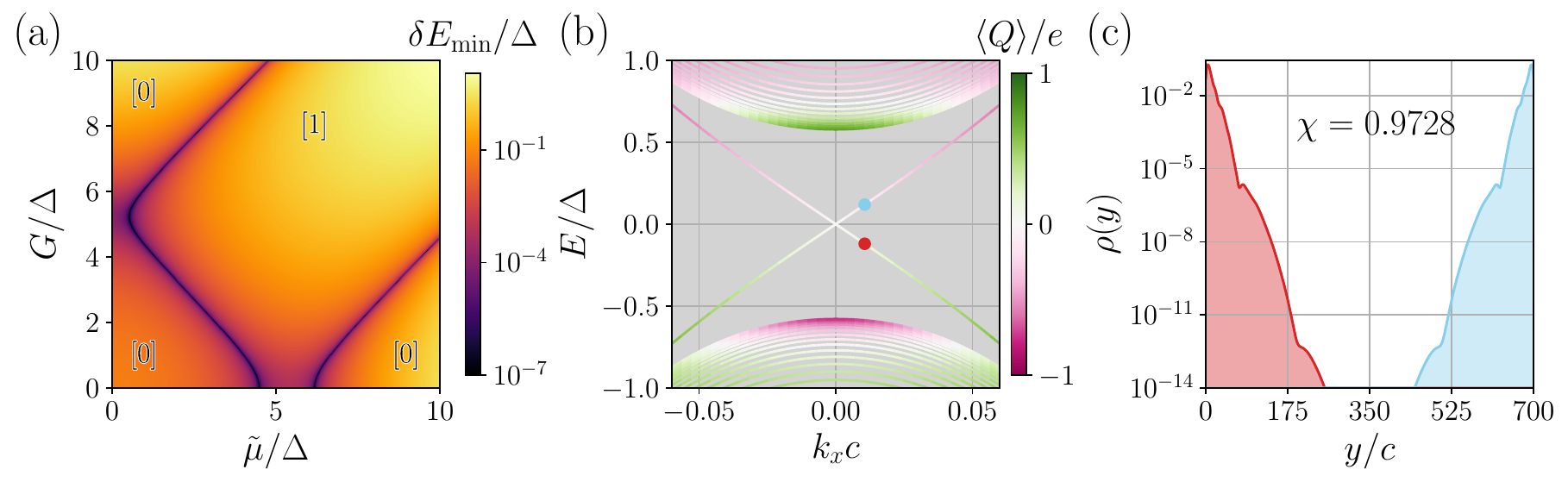}
    \caption{
\textbf{Bulk phase diagram and chiral Majorana edge modes of the superconducting \gls{bhz} ribbon.}
\textbf{(a)} Bulk phase map in the dimensionless $(\tilde{\mu}/\Delta,G/\Delta)$ plane at fixed inversion-breaking term \(T/\Delta=4.5\). The colour scale shows the smallest absolute BdG eigenenergy \(\delta E_{\min}/\Delta\), obtained by numerical diagonalization, while the black contours are the analytical gap-closing boundaries of \cref{eq:GapClosings}. The labels $[|C|]$ indicate the absolute value of the region Chern number.
\textbf{(b)} Spectrum of a ribbon periodic along $\hat{x}$ and finite along $\hat{y}$, with width \(L_y/c=700\) for fixed $(\tilde{\mu}/\Delta,G/\Delta)= (1.0, 4.6)$. The bands are coloured by the normalized charge expectation \(\langle Q\rangle/e\) defined in \cref{eq:charge-expectation}.
\textbf{(c)} Probability density $\rho(y)$
of the two highlighted states  with \(k_xc\simeq0.0106\) of panel~\textbf{(b)}. The states are localized on opposite ribbon edges and have Majorana polarization $\chi\simeq0.97$.
}
\label{Fig2}
\end{figure}

\Cref{Fig2}\textbf{(a)} shows the resulting phase diagram in the \((\tilde{\mu},G)\) plane. In the parameter window relevant for the rest of the work, the gap closings separate a trivial superconducting phase from a chiral phase with Chern number \(C=-1\). To verify the boundary physics, we diagonalize a ribbon that is periodic along \(\hat{x}\) and open along \(\hat{y}\). The spectrum in \cref{Fig2}\textbf{(b)} shows a single pair of counter-propagating low-energy branches crossing the bulk gap, as expected for a class-D chiral superconductor with \(|C|=1\). As shown in \cref{Fig2}\textbf{(c)}, the linearly dispersive states have nearly unit Majorana polarization, \(\chi\simeq0.97\) for \(k_xc=0^+\), and are localized at opposite edges: left (right) edge for the negative (positive) velocity branch.
The phase maps are computed using the Bi$_2$Se$_3$-motivated dimensionless parameters introduced in~\cref{sec:ModelAndDiagnostics} for \(T/\Delta=4.5\).

We thus establish the two-dimensional \(C=-1\) phase as the basic unit for the multi-channel construction.
In the following sections this phase is used to obtain high Chern number states via the inclusion of a structure in the $\hat{z}$ direction.
Related chiral superconducting phases and Majorana edge states in \gls{bhz}-type systems have also been analyzed in other two-dimensional contexts, including transition-metal dichalcogenide platforms and unconventional pairing structures~\cite{Qi2010,Novik2020,Wang2014, Ji2024}.

\section{Proximitized three-dimensional slab in the quantum well regime}
\label{sec:Stacking}

We now use the two-dimensional class-D phase of \cref{sec:2Dlimit} as a building block, extending it along the $\hat z$ axis in order to engineer multi-channel chiral-Majorana phases.
In a first approximation, we consider the proximity effect to be uniform, i.e., $\Delta(z) = \Delta$ with $\Delta$ constant,
and that the localization in the vertical direction creates multiple independent subbands.
Analogously to high-Chern states in non-superconducting magnetic systems~\cite{Wang2013, Wang2021, Baba2022}, each subband in the proximitized case may contribute to the formation of a superconducting topological mode. These modes then host chiral Majorana states in a finite system with an effective Majorana number \(\left|C\right|>1\).

\par To capture the simplest \gls{3d} effects while preserving the algebra of \cref{sec:2Dlimit}, we include only the quadratic dispersion in $\hat{z}$. The mass term is then given by the full \cref{eq:MassFull3D}, with a non-zero term proportional to $M_1$, but we still keep $C_1=B=0$ in \cref{eq:FullBdG}.
Since considering $M_1 \neq 0$ merely augments the mass term, all anticommutation relations among the Dirac matrices remain intact. Thus, the analytic gap closing condition in~\cref{eq:GapClosings} and the bulk-invariant calculations of Appendix~\ref{app:2d-derivation} remain valid, but now depend on $k_z$.

For a finite slab of thickness $L_z=Nc$, with $N$ the number of vertical layers, the momentum dependence along $\hat{z}$ enters only through the quadratic mass term. Consequently, the spectrum consists of quantum-well subbands labeled by the quantized momenta $k_z^{(m)}c=\pi m/(N+1)$.
This quantization directly links the bulk topology to the confined geometry,
enabling the three-dimensional topological phase diagram to be inferred from the corresponding two-dimensional one.
In practice, we first consider $k_z$ as a parameter and calculate the Chern number $C(k_z)$ for the corresponding two-dimensional system. Then, we identify the critical $k_z^{\pm}$ values at which the bulk gap closes and count how many $k_z^{(m)}$ modes remain topological for a given ribbon thickness \(L_z\). Whenever \(\sum_{m = 1}^N C(k_z^{(m)})=\pm n\neq0\), the slab hosts \(\left|n\right|\) co-propagating \glspl{mzm} at each edge.
More compactly, the Chern number is determined by the counting rule
\begin{equation} \label{eq:QW_Csum}
C_{\text{slab}}=\sum_{m=1}^{N}C\!\left(k_z^{(m)}\right)\Theta\bigl(k_z^{(m)}-k_z^{-}\bigr)\,\Theta\bigl(k_z^{+}-k_z^{(m)}\bigr),
\end{equation}
where \(\Theta\) is the Heaviside function and $k^{\pm}_z$ are the values for which a topological transition occurs for some fixed set of parameters \(\{G,\Delta,T,\tilde{\mu}\}\).
The factor $C(k_z^{(m)})$ retains the sign of each two-dimensional contribution; in the topological window considered below, every subband contributes with $C(k_z^{(m)})=-1$.

\begin{figure}[t]
    \centering
    \includegraphics[width=\linewidth]{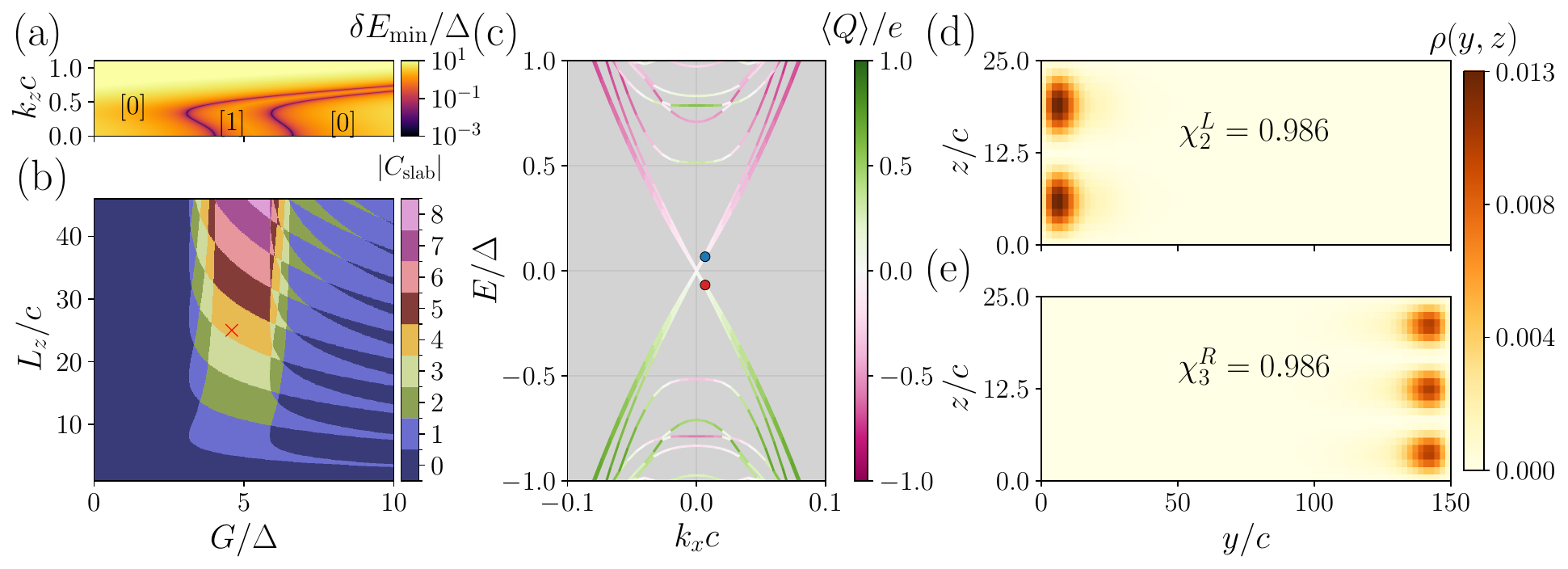}
\caption{
\textbf{High-Chern-number superconducting slab and representative Majorana modes.}
\textbf{(a, b)} Phase diagrams for the vertically stacked $\mathcal H_{\rm BdG}(\mathbf k)$ with \(B = 0\),~\cref{eq:FullBdG}, as a function of the dimensionless Zeeman field \(G/\Delta\) at fixed \(T/\Delta=4.5\).
\textbf{(a)} Minimum bulk gap \(\delta E_{\min}(k_z,G)/\Delta\) as a function of the momentum \(k_zc\) for a system periodic in all spatial directions, in a logarithmic color scale where dark contours indicate bulk gap closings.
\textbf{(b)} Total Chern number \(|C_{\mathrm{slab}}|\) for a slab with finite thickness \(L_z/c\).
The red cross marks the parameters \(G/\Delta=4.6\) and \(L_z/c=25\) used in panels \textbf{(c-e)}.
\textbf{(c)} Low-energy spectrum of the corresponding slab periodic along the $\hat x$ direction and finite along $\hat y$ and $\hat z$, with \(L_y/c=150\) and \(L_z/c=25\). The colour scale gives the normalized charge expectation \(\langle Q\rangle/e\) defined in~\cref{eq:charge-expectation}. The colored markers identify the two representative states.
\textbf{(d, e)} Probability density $\rho(y,z)$ of the selected states at
\(k_xc\simeq0.0067\). The labels indicate the edge localization
($L$ or $R$), the quantum-well mode index $m$, and the corresponding Majorana polarization $\chi_m^{L,R}$.
}
\label{Fig3}
\end{figure}

The interplay between bulk topology and quantum confinement is illustrated in~\cref{Fig3}. Panel~\textbf{(a)} shows the dimensionless minimum gap \(\delta E_{\min}/\Delta\) as a function of the momentum \(k_z\) and Zeeman splitting \(G\), revealing a nontrivial topological region with \(C = -1\) for magnetization \(G/\Delta \gtrsim 3\) and for intermediate values of $k_zc$.
While we consider periodic boundary conditions along the vertical direction in \cref{Fig3}\textbf{(a)}, the resulting topological phases as a function of $k_z$ map onto the phase diagram of a slab with finite thickness $L_z$ [\cref{Fig3}\textbf{(b)}], where we counted the number of quantum-well modes satisfying condition~\eqref{eq:QW_Csum}.
In the finite slab case, the phase diagram replicates the topological regions of the $k_z$-dependent map in \cref{Fig3}\textbf{(a)} in a staircase pattern
since~\cref{eq:QW_Csum} dictates a correspondence between the \textit{discrete} $k_z^{(m)}$ values in a slab and the non-trivial $C(k_z)$, with $k_z$ a \textit{parameter} for the two-dimensional calculation of the Chern number.
Both the slab thickness and the parameter \(G\) thus offer independent knobs to control the multiplicity of chiral \glspl{mzm}.

\par We now analyze the slab configuration corresponding to the red cross in Fig.~\ref{Fig3}\textbf{(b)}, which is located in a regime with a total Chern number \(|C_{\text{slab}}|=4\). The \gls{bdg} spectrum [Fig.~\ref{Fig3}\textbf{(c)}] exhibits eight linearly dispersing subgap branches, split equally between right- and left-movers, realizing four chiral Majorana channels per edge.
The dispersion relation in \cref{Fig3}\textbf{(c)} also displays in a colour map the projection of the expectation value of the charge, \cref{eq:charge-expectation}. As expected, the subgap states have \(|Q_\nu|\ll e\). The Majorana polarization, \cref{eq:regional-majorana-polarization}, yields consistent results: the quantum-well subgap modes satisfy \(\chi_m^{L,R}\simeq 1\) for $m = 1, \dots, 4$, confirming the states self conjugation. We have also verified that the Majorana polarization effectively vanishes for bulk excitations.
\Cref{Fig3}\textbf{(d,e)} show the local density $\rho$ and the Majorana polarization \(\chi_m^{L,R}\) for the two chiral Majorana modes with quantum-well index $m = 2,3$ and localized on the left ($L$) and right ($R$) sides. At small positive momentum $k_x$, the chiral states have negative (positive) energy for the left (right) localization.
The local density $\rho$ shows the real-space structure typical of a quantum well along \(\hat{z}\) featuring $m$ maxima, which directly reflects the transverse mode index.
The subgap states thus act as independent confined modes of a slab and each chiral \gls{mzm} develops a standing-wave envelope across the thickness $\hat{z}$, together with a well-localized profile in the in-plane directions.
The strong decay of the probability density in the $\hat{y}$ direction in \cref{Fig3}\textbf{(d,e)} further corroborates the topological nature of the previously identified edge states, confirming their interpretation as well-separated chiral \gls{mzm}.

\section{Proximitized slab with linear Dirac coupling in the vertical direction}
\label{sec:LinearZ}

\begin{figure}[t]
    \centering
    \includegraphics[width=\linewidth]{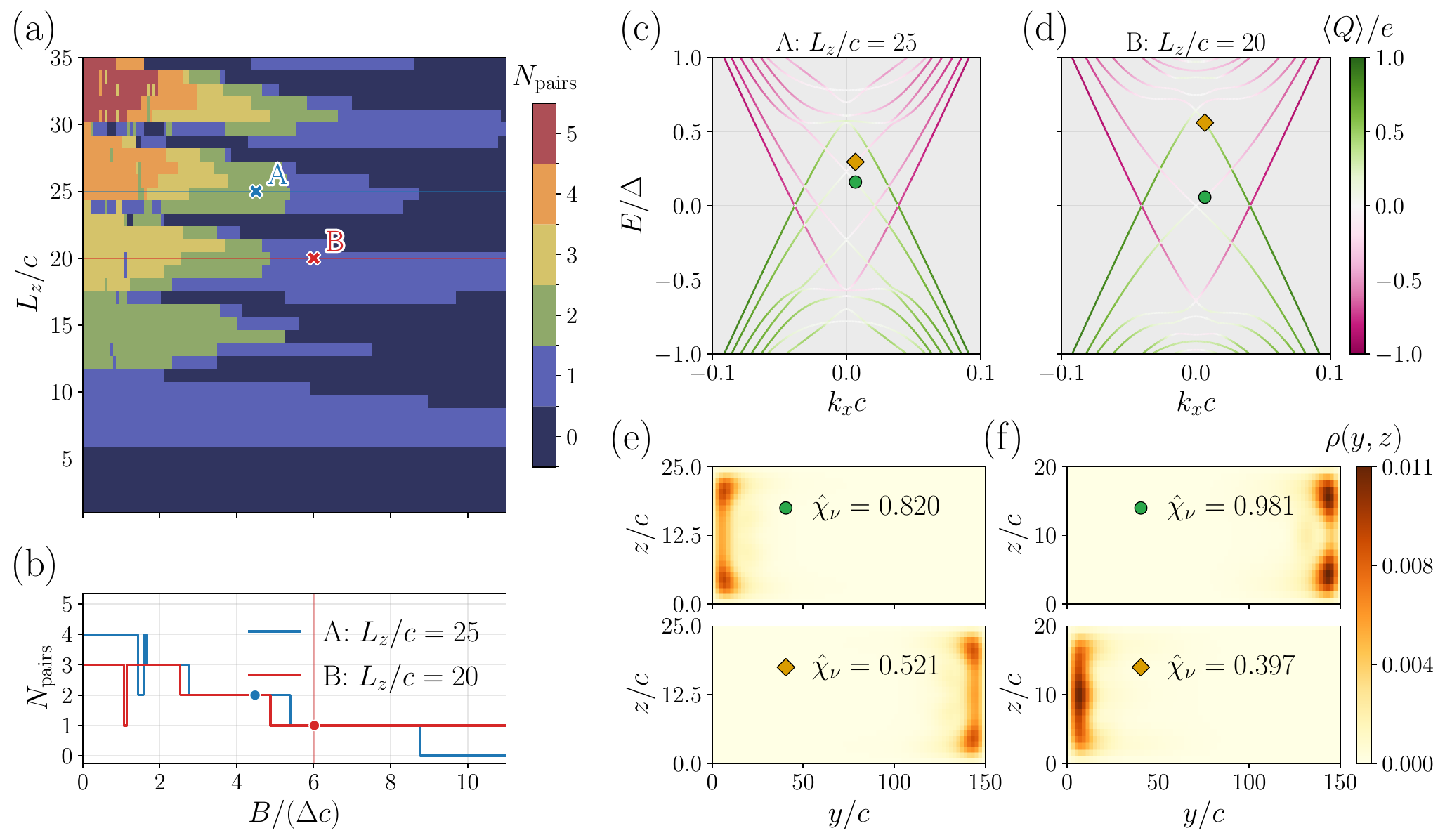}
\caption{
\textbf{Effect of the linear $k_z$ term on the multiplicity of Majorana pairs.}
\textbf{(a)} Numerical pair-count map as a function of slab thickness $L_z/c$ and linear coefficient $B/(\Delta c)$, for fixed $L_y/c=150$, $T/\Delta=4.5$ and $|G|/\Delta=4.6$.
The colour scale represents $N_{\mathrm{pairs}}$, obtained from the low-energy states at $k_x c\simeq6.7\times10^{-3}$.
A state contributes to the count when its Majorana polarization satisfies $\hat{\chi}_\nu>0.5$ and its charge expectation value obeys $|\langle Q\rangle|<0.1e$. The blue and red crosses mark the representative points A and B.
\textbf{(b)} Cuts of the map in panel~\textbf{(a)} at the two selected slab thicknesses. Point A corresponds to $B/(\Delta c)=4.5$, $L_z/c=25$ and $N_{\mathrm{pairs}}=2$, whereas point B corresponds to $B/(\Delta c)=6$, $L_z/c=20$ and $N_{\mathrm{pairs}}=1$.
\textbf{(c, d)} Low-energy spectra \(E/\Delta\) versus \(k_xc\) for points A \textbf{(c)} and B \textbf{(d)} in a slab finite along \(\hat y\) and \(\hat z\) and periodic along \(\hat x\), coloured by the normalized charge expectation \(\langle Q\rangle/e\).
The green circles and gold diamonds respectively identify the lowest and next-lowest positive-energy bands at the first positive momentum \(k_xc\simeq6.7\times10^{-3}\) used for the pair count.
At point A we have \((E/\Delta,\hat{\chi}_\nu)\) equal to
\((0.161,0.820)\) and \((0.296,0.521)\) and at point B,
\((0.0584,0.981)\) and \((0.561,0.397)\).
\textbf{(e, f)} Probability densities \(\rho(y,z)\) of the marked states for points A \textbf{(e)} and B \textbf{(f)}, with the green states on top and gold ones below.
The corresponding Majorana polarization \(\hat{\chi}_\nu\) is also indicated.
}
    \label{Fig4}
\end{figure}

In \cref{sec:Stacking}, we analyzed the quantum-well limit with $B=0$, which allowed us to assign an approximate quantized transverse momentum $k_z^{(m)}$ to the confined states along $\hat z$ and to treat them as independent two-dimensional modes.
While this limit captures the mechanism behind the multi-channel phase, it omits a key ingredient of the three-dimensional \gls{ti} Hamiltonian: the linear, odd-in-$k_z$ coupling between opposite orbital-parity sectors~\cite{Zhang2009,Wang2013}.
We now include this effect by setting $B\neq0$ in the full Hamiltonian, \eqref{eq:FullBdG}.
In a slab with open boundaries along $\hat z$, the standing-wave subbands are formed from superpositions of $\pm k_z$ components; the $B$ term in~\cref{eq:FullBdG} therefore hybridizes the quantum-well modes and invalidates the calculation of the Chern-number based on the counting rule defined in~\cref{eq:QW_Csum}.
Consequently, in the case of the full Hamiltonian with the linear term, the number of chiral Majorana states has to be computed directly in the finite slabs using numerical criteria based on the Majorana polarization and charge. Specifically, since the quantum-well index \(m\) is no longer a suitable label we instead use the edge-resolved indicator $\hat{\chi}_\nu$, which is defined for a given numerical wavefunction labeled by $\nu$ as the maxima of the two edges
\begin{equation}
    \hat{\chi}_\nu
    \equiv
    \max\!\left(\chi_\nu^L,\chi_\nu^R\right).
    \label{eq:edge-resolved-majorana-indicator}
\end{equation}
Moreover, to avoid the numerical degeneracies occurring at zero momentum we evaluate the states at a small positive value of \(k_xc=0^+\).
See Appendix~\ref{app:linear-kx0} for the specific Majorana pair count at the particle-hole invariant point $k_x=0$, which clarifies this issue of numerical degeneracies in the counting method.

The evolution of the Chern phases as a function of the linear term $B$ and the slab width $L_z$ is shown in \cref{Fig4}\textbf{(a)}. The Chern states are defined from a numerical calculation of the number of \gls{mzm} pairs, denoted by $N_{\rm pairs}$, obtained from the full finite-slab spectrum after using the Majorana polarization and charge criteria as described in the figure caption.
In the phase map of \cref{Fig4}\textbf{(a)} the staircase structure found in the $B=0$ quantum-well limit survives roughly up to $B/(\Delta c)\gtrsim 4$ values of the linear coupling.
Further increasing the linear coupling $B$ reduces the number of \glspl{mzm} in steps of two:
For the $L_z/c=25$ cut (blue line), panel~\textbf{(b)} shows how $N_{\rm pairs}$ evolves from four pairs at $B=0$ to two pairs near point A and eventually to no counted pairs at large
$B$.
Similarly, for the $L_z/c=20$ cut (red line), the count is reduced
from three pairs to a single remaining pair near point B.
Note that the phase map in \cref{Fig4}\textbf{(a)} shows that the cases with an odd number of $N_{\rm pairs}$ in the quantum-well limit, i.e., when $B=0$, exhibit a more robust nontrivial topological phase at high $B$. As we explain next, such robustness comes from a particular coupling scheme of the high-Chern chiral \glspl{mzm}.

The hybridization between high-Chern \glspl{mzm} can be understood by comparing the spectra and local densities at $B=0$ shown in \cref{Fig3}\textbf{(c-e)} with the $B\neq0$ cases in \cref{Fig4}\textbf{(c-f)}.
At $B=0$, all Majorana edge states display a linear dispersion crossing zero energy at $k_xc=0$. By contrast, at finite values of the linear term $B$ only the cases with odd number of $N_{\rm pairs}$ at $B=0$ feature one remaining band with zero-energy crossings at $k_xc=0$.
For example, at point A the system still contains two Majorana-like pairs, but the ideal quantum-well structure of~\cref{Fig3} has already degraded.
As shown in the spectra of \cref{Fig4}\textbf{(c)}, for slab configurations with an odd number of $N_{\rm pairs}$ at $B = 0$, the linear term shifts the Dirac-like cone of the chiral bands to higher energies. Consequently, the selected low-energy states at point A have reduced Majorana character.
Indeed, the two lowest positive-energy states displayed in panel \textbf{(e)} remain edge-localized, but their edge-resolved Majorana indicators are reduced to \(\hat{\chi}_\nu\simeq0.820\) and \(0.521\).

On the other hand, at point B a linear chiral band is still centered at zero energy and momentum, see panel \textbf{(d)}, and one pair of zero-energy modes still satisfies the selection criteria leading to $N_{\rm pairs} = 1$. The near-zero state shown at the top of panel \textbf{(f)} has \(\hat{\chi}_\nu\simeq0.981\), whereas the next positive-energy band, shown at the bottom, has
\(\hat{\chi}_\nu\simeq0.397\) and fails the selection criteria. Such a contrast directly indicates how a single chiral Majorana channel survives after the additional subbands have hybridized away.

For the family of ${\rm Bi_2Se_3}$, ${ \rm Bi_2Te_3}$, and ${\rm Sb_2Te_3}$, the linear term spans $B/(\Delta c) \sim 1.5-11.5$ ~\cite{Liu2010}, i.e., $B \sim 0.30 - 2.3~\si{eV~\angstrom}$ for $\Delta c=0.2~\si{eV~\angstrom}$.
Therefore, the linear vertical coupling is indeed a non-negligible term in these materials and sets a practical upper limit on the multiplicity of clean quantum-well Majorana channels.
While it does not immediately destroy all chiral \glspl{mzm}, it progressively removes the independent-subband structure on
which the high-multiplicity phases rely. Tuning either the slab thickness $L_z$ or the strength of this orbital-mixing term can thus drive the finite system between multi-pair, single-pair, and fully hybridized regimes.

\section{Resilience of the multiple chiral Majorana phase to disorder}\label{sec:disorder}

\begin{figure}[t]
    \centering
    \includegraphics[width=\linewidth]{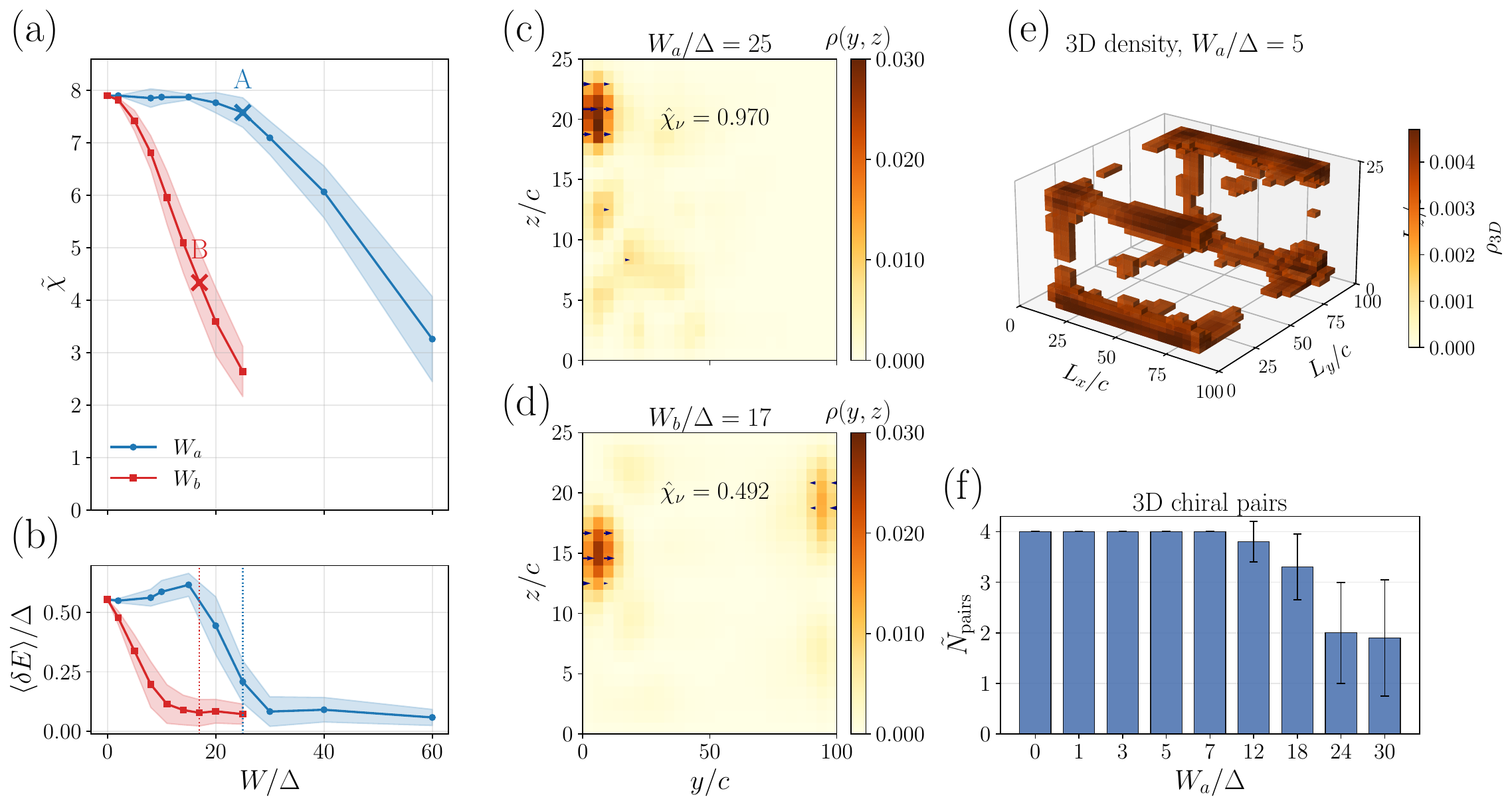}
    \caption{
\textbf{Disorder response of chiral Majorana modes in a semi-infinite and fully finite slab.}
\textbf{(a)} Disorder evolution of the total edge Majorana polarization $\tilde{\chi}$, \cref{eq:majorana-pol-av}, for the two disorder channels $\delta \mathcal{H}_a$ and $\delta \mathcal{H}_b$ in \cref{eq:disorder}.
Solid lines show the mean value over 30 disorder realizations and shaded regions indicate the standard deviation. We highlight the points A with \(W_a/\Delta=25\) and B with \(W_b/\Delta=17\).
\textbf{(b)} Evolution of the finite-size low-energy gap proxy \(\langle\delta E\rangle/\Delta\) (the fifth-smallest absolute energy near zero) as a function of disorder amplitude.
\textbf{(c, d)} Probability density \(\rho(y,z)\) of the selected low-energy states for points A and B, respectively, in a slab periodic along \(\hat x\) and finite along \(\hat y\) and \(\hat z\), with \(L_y/c=100\) and \(L_z/c=25\).
The states are calculated at \(k_xc=0^+\). The black arrows show the local Majorana polarization, $\mathbf{MP}_\nu(\vec{r})$, with its real and imaginary components superimposed on the density.
The selected states have edge-resolved indicators \(\hat{\chi}_\nu=0.970\) for A and \(\hat{\chi}_\nu=0.492\) for B.
\textbf{(e)} \gls{3d} low-energy density $\rho_{3D}$ for electrostatic disorder \(W_a/\Delta=5\) in a finite system with \(L_x/c=L_y/c=100\) and \(L_z/c=25\). The voxel plot displays the dominant low-energy probability density.
\textbf{(f)} Averaged number $\tilde{N}_{\rm pairs}$ of pairs of \gls{3d} chiral modes identified from the winding of the vector Majorana polarization $\gamma$, \cref{eq:majorana-polarization-winding}, as a function of electrostatic disorder strength $W_a$.
States are counted when the winding number \(\gamma\) satisfies \(|\gamma|=1\pm0.25\) for \(|E|/\Delta<1\), and the result is divided by two to obtain the pair count.
}
\label{Fig5}
\end{figure}

We now test the stability of the multiple chiral \glspl{mzm} against disorder.
To do this, we consider two different random local perturbations, defined as
\begin{equation} \label{eq:disorder}
    \delta\mathcal H_a(\mathbf r)
    =
    W_a \eta_a(\mathbf r) \hGamma{z00},
    \qquad
    \delta\mathcal H_b(\mathbf r)
    =
    W_b \eta_b(\mathbf r)\hGamma{0yy}~.
\end{equation}
Here \(W_\alpha\) is the disorder strength and \(\eta_\alpha(\mathbf r)\in[-1/2,1/2]\), with $\alpha = a,b$, is a uniform random number uncorrelated from site to site.
The \(W_a\) channel is ordinary electrostatic Anderson disorder, or equivalently, a random local chemical potential in the \gls{bdg} Hamiltonian.
The \(W_b\) channel is different: it represents a random onsite perturbation with the same internal matrix structure as the linear \(k_z\) coupling in \cref{eq:FullBdG} and, therefore, directly breaks the mode-decoupling symmetry of the quantum-well limit.
Because this onsite term lacks the odd-in-$k_z$ factor of the clean linear coupling, it also explicitly breaks the particle-hole constraint of the \gls{bdg} Hamiltonian. We use $W_b$ only as a symmetry-breaking control channel, rather than as a particle-hole-symmetric disorder model, to contrast its response with the particle-hole-symmetric electrostatic disorder $W_a$.
In the following, we express the disorder amplitudes in units of  \(\Delta\) and perform simulations both in the semi-infinite geometry, i.e., with $k_x$ a well-defined momentum, and in a fully finite system in the three spatial dimensions.

\par For concreteness, we study the resilience to disorder of a slab with \(L_z/c=25\) and \(B = 0\).  In the absence of disorder, this corresponds to a system with clearly defined quantum-well modes and \(C_{\rm slab} = -4\).
First, for a semi-infinite system with well-defined $k_x$, we diagonalize the finite \((y,z)\) slab at \(k_x=0^+\).
Since eight chiral states are expected for $C_{\rm slab} = -4$, we select the eight states closest to zero energy for each disorder realization and compute the corresponding \(\hat{\chi}_\nu\) according to \cref{eq:edge-resolved-majorana-indicator}.
To track the resilience to disorder, in \cref{Fig5}\textbf{(a)} we analyze the averaged total Majorana polarization defined as:
\begin{equation} \label{eq:majorana-pol-av}
   \tilde{\chi}= \left \langle \sum_{\nu\in\mathcal S}\hat{\chi}_\nu \right \rangle_{\rm avg} ,
\end{equation}
with $\mathcal S$ indicating the set of $2|C_{\rm slab}|$ states with lowest absolute energy (eight states for $C_{\rm slab} = -4$) and where the average is taken over disorder realizations.
In the clean limit, since all chiral modes have $\chi_\nu \sim 1$, the total averaged value is $\tilde{\chi} \approx 2 |C_{\rm slab}|$.
Therefore, the starting point at $W_{a,b}=0$ in \cref{Fig5}\textbf{(a)} is close to eight with almost no deviation.
In \cref{Fig5}\textbf{(b)} we complement the averaged Majorana polarization with a plot of the averaged bulk gap \(\langle\delta E\rangle\) of the slab, defined as the disorder average of the \((|C_{\rm slab}|+1)\)-th smallest absolute energy, in this case the fifth one.
Finally, the spatial maps of \cref{Fig5}\textbf{(c,d)} show, for one disorder realization, the probability density (colours) and the local Majorana polarization [arrows, see~\cref{eq:local-majorana-polarization}].

\par Disorder is expected to decrease the bulk gap until it eventually destroys the topological protection, thus impacting the Majorana character of the low energy states.
However, as shown in \cref{Fig5}\textbf{(a-b)}, the response to disorder for both the averaged Majorana polarization $\tilde{\chi} $ and average gap size $\langle\delta E\rangle$ is strongly dependent on the symmetry of the perturbation term in \cref{eq:disorder}.
Diagonal electrostatic disorder $\delta \mathcal{H}_a$ leaves the Majorana polarization almost unchanged up to point A. At \(W_a/\Delta=25\), the total average polarization is  \(\tilde{\chi}\simeq7.6\) and a representative edge state plotted in \cref{Fig5}\textbf{(c)} still has \(\hat{\chi}_\nu=0.970\), although the gap in panel~\textbf{(b)} is already reduced to less than half the value without disorder.
By contrast, the symmetry-breaking channel $\delta \mathcal{H}_b$ suppresses the Majorana character much faster. At point B with \(W_b/\Delta=17\), the total polarization has fallen to \(\tilde{\chi}\simeq4.3\), the bulk gap is almost closed, and the
representative state in \cref{Fig5}\textbf{(d)} has only \(\hat{\chi}_\nu=0.492\).
The spatial maps further illustrate this point:
For \(W_a\) the local density and Majorana-polarization weight remain concentrated on a single edge, while for \(W_b\) the state has appreciable weight on both sides of the slab and a less localized Majorana polarization.

We also examined a fully finite \gls{3d} system under electrostatic
disorder. The calculation uses a finite slab with \(L_x/c=L_y/c=100\) and \(L_z/c=25\), with random disorder $\delta \mathcal{H}_a$ on every site. For each low-energy eigenstate, we evaluate the phase winding of the local Majorana polarization vector along the closed boundary contour introduced in \cref{eq:majorana-polarization-winding};
in the discrete calculation the contour is followed at each \(z\) layer and the resulting winding is averaged over \(z\).
We then count states with \(|\gamma|=1\pm0.25\) and \(|E|/\Delta<1\) and define $\tilde{N}_{\rm pairs}$ as half the number of those states.
The clean value \(\tilde{N}_{\rm pairs}=4\) is preserved throughout the weak-disorder regime and remains unchanged up to \(W_a/\Delta\simeq7\); see \cref{Fig5}\textbf{(f)}. The density map at \(W_a/\Delta=5\) in \cref{Fig5}\textbf{(e)} confirms that the low-energy probability density remains concentrated on the boundary of the finite sample.
Therefore, electrostatic Anderson disorder  does not immediately destroy the chiral \glspl{mzm} even in fully finite structures where the finite-size effects could increase the effective coupling.
By contrast, perturbations that break the symmetry of the quantum-well modes such as $\delta \mathcal{H}_b$ have a much greater effect on the topological modes.

\section{Tuning the multiple Majorana phase in a Josephson junction}\label{sec:JJ}

\begin{figure}[t]
    \centering
    \includegraphics[width=\linewidth]{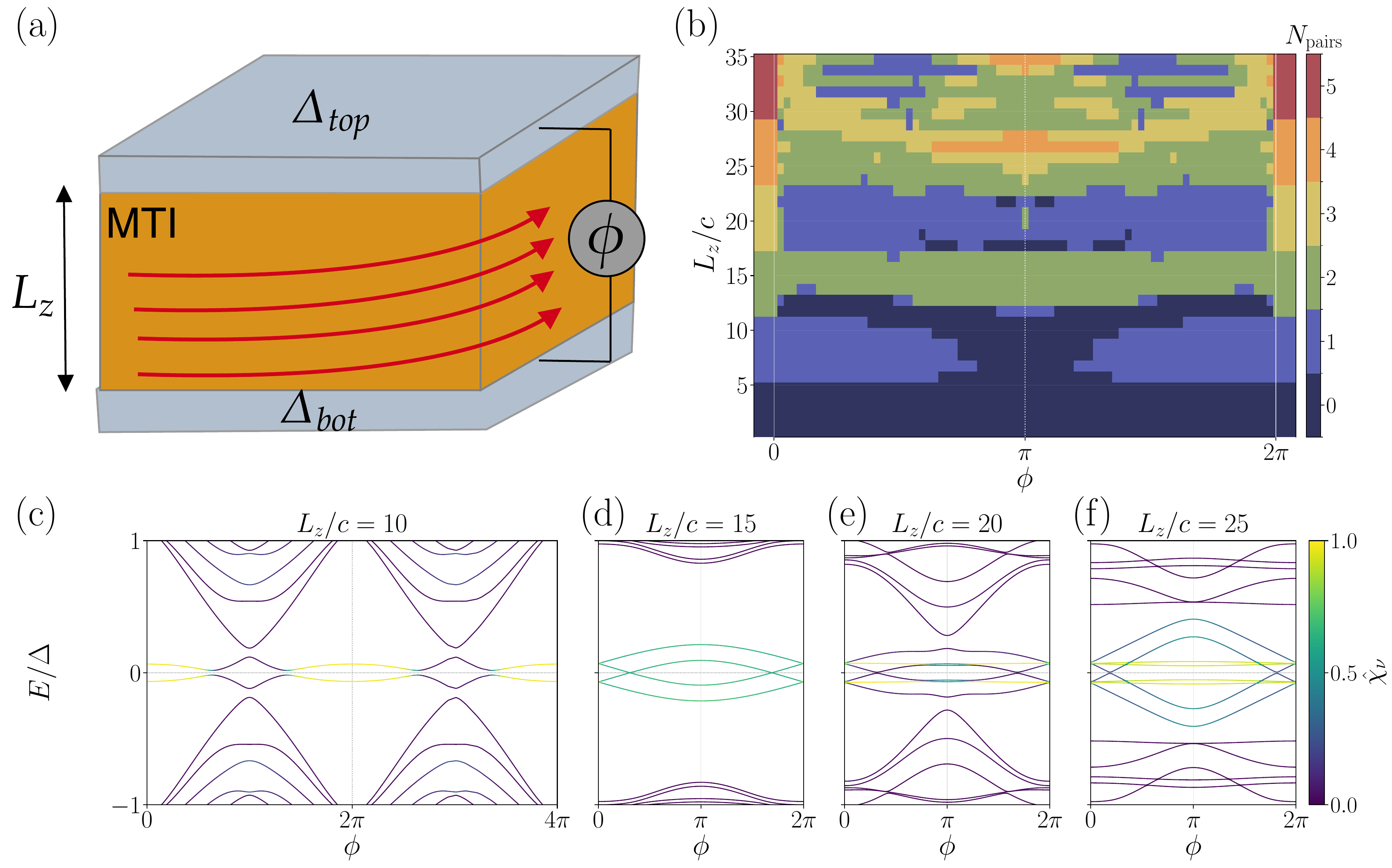}
    \caption{
\textbf{Josephson phase control of Majorana-pair multiplicity.}
\textbf{(a)} Schematic of the vertical Josephson geometry. The induced pairing is described by \cref{eq:deltaJJ} with a relative phase \(\phi\) between the top and the bottom half of the slab.
\textbf{(b)} Pair-count map versus the dimensionless slab thickness \(L_z/c\) and phase difference \(\phi\) for fixed \(L_y/c=150\), \(T/\Delta=4.5\), and \(|G|/\Delta=4.6\).
The colour scale indicates \(N_{\mathrm{pairs}}\) computed from the twenty states closest to zero energy at \(k_xc\simeq0.0067\). A state is selected when \(\hat{\chi}_\nu>0.5\), and the pair count is half the number of selected states.
The unlabelled strips outside the white boundary lines in the horizontal axis repeat the \(\phi=0\) and \(2\pi\) results for visualization only.
\textbf{(c-f)} Energy spectra versus \(\phi\) for \(L_z/c=10,15,20,\) and \(25\) at the same momentum.
Panel \textbf{(c)} displays \(0\leq\phi\leq4\pi\), whereas panels \textbf{(d-f)} display one period, \(0\leq\phi\leq2\pi\).
The bands are coloured by the edge-resolved Majorana polarization \(\hat{\chi}_\nu\) in \cref{eq:edge-resolved-majorana-indicator}.
}
\label{Fig6}
\end{figure}

As a last tuning knob of the multiple chiral Majorana phase, we consider a vertical Josephson geometry where the slab is proximitized by two superconducting regions with a tunable relative phase difference \(\phi\); see schematics in \cref{Fig6}\textbf{(a)}. In the calculation, the pairing amplitude has the same magnitude in the two halves of the system but acquires a phase jump across the middle of the slab,
\begin{equation} \label{eq:deltaJJ}
\Delta(z)=
\begin{cases}
\Delta e^{i\phi_{\mathrm{top}}}, & z>L_z/2,\\
\Delta e^{i\phi_{\mathrm{bot}}}, & z<L_z/2,
\end{cases}
\end{equation}
with \(\phi_{\mathrm{top,bot}}\) the phase of the top and bottom superconductors and $\phi = \phi_{\mathrm{bot}} - \phi_{\mathrm{top}}$.
For this complex phase profile, the pairing contribution is implemented as \(\mathcal H_{\Delta}(z)=\operatorname{Re}[\Delta(z)]\,\hGamma{yyz}+\operatorname{Im}[\Delta(z)]\,\hGamma{xyz}\), so that the two matrices carry real coefficients and the full \gls{bdg} Hamiltonian remains Hermitian. The pairing term in \cref{eq:FullBdG} is the corresponding shorthand for a real order parameter.
For simplicity, in this section we only consider the limiting case of the quantum-well basis, i.e., with $B = 0$. This way, the phase difference provides a controllable way of hybridizing the Majorana channels associated with different quantum-well modes,
while keeping the same normal state confinement physics discussed above.

The phase diagram as a function of the slab thickness \(L_z\) and relative phase \(\phi\) is shown in \cref{Fig6}\textbf{(b)}. The different topological phases are characterized by the number of Majorana-like pairs $N_{\rm pairs}$ obtained after numerical diagonalization of the semi-infinite slab at $k_x c = 0^+$ by counting the zero-energy modes that fulfill the conditions discussed in \cref{sec:LinearZ}.
At zero Josephson phase (and multiples of $2\pi$) the system recovers the quantum-well counting of \cref{sec:Stacking} where increasing \(L_z\) introduces additional quantum-well subbands and, therefore, increases the number of chiral Majorana pairs.
Conversely, increasing the Josephson phase $\phi$ can substantially change the number of states that fulfill the Majorana polarization criterion, most pronouncedly around \(\phi=\pi\).
As we argue later, at this phase the sign reversal of the induced order parameter between the two halves of the slab enhances the coupling between modes that are separated by their transverse structure.
Note that the map is \(2\pi\)-periodic in $\phi$, as expected for the static \gls{bdg} spectrum.
A gauge-invariant check of the same Josephson geometry at \(k_x=0\), including the finite-size gap scaling and the absence of a protected \(4\pi\) crossing, is collected in Appendix~\ref{app:4pi}.

\par The spectra in~\cref{Fig6}\textbf{(c-f)} illustrate how this phase-controlled hybridization depends on the parity and multiplicity of the available channels.
For \(L_z/c=10\) [\cref{Fig6}\textbf{(c)}], the single Majorana pair present at \(\phi=0\) loses its Majorana character around \(\phi=\pi\), indicating that an isolated pair can be fully quenched by the Josephson coupling.
By contrast, the \(L_z/c=15\) case [\cref{Fig6}\textbf{(d)}] contains two pairs with reduced but still notable Majorana polarization that remain visible throughout the cycle. For thicker slabs, where three or four pairs are present at \(\phi=0\), the phase difference removes some but not all of the low-energy Majorana-like modes, consistently with a pairwise hybridization mechanism; see the horizontal stripe-like pattern in the phase map in \cref{Fig6}\textbf{(b)}.

\begin{figure}[h]
    \centering
    \includegraphics[width=\linewidth]{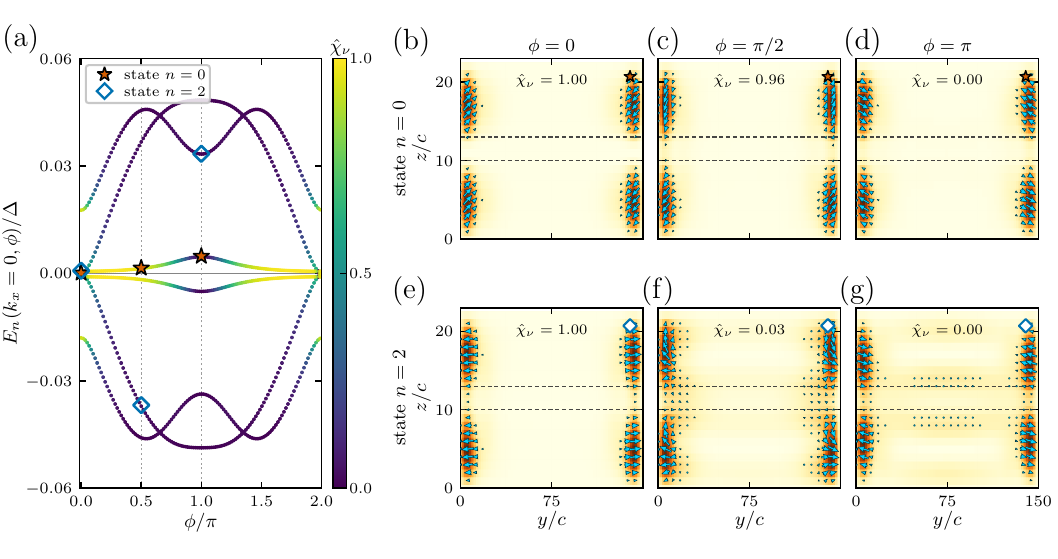}
    \caption{
    \textbf{Phase-driven loss of Majorana coherence in the insulating-barrier junction.}
    \textbf{(a)} Low-energy spectrum of the \(S(10)/I(3)/S(10)\) structure at \(k_x=0\), with every band coloured by the edge-resolved Majorana indicator \(\hat{\chi}_\nu\). Orange stars and blue diamonds mark, respectively, the \(n=0\) and \(n=2\) states shown in the spatial panels at \(\phi=0,\pi/2,\pi\).
    \textbf{(b-g)} Normalized probability density \(\rho_n(y,z)/\rho_{n,\max}\) with the local Majorana-polarization vector $\textbf{MP}(\vec{r})$
    overlaid for \(n=0\) [panels \textbf{(b-d)}] and \(n=2\) [panels \textbf{(e-g)}]. Arrow lengths are normalized independently in every panel, and cyan arrows are shown only where the density is significant.
    When \(\hat{\chi}_\nu\simeq1\) the vectors within each edge region are aligned and add coherently. As \(\phi\to\pi\), they rotate out of phase and cancel region by region, even though the states remain predominantly edge-localized.}
    \label{Fig7}
\end{figure}

By considering the case of an insulating barrier between the two proximitized regions we gain a deeper intuition of how the hybridization emerges across the Josephson junction and why it is more pronounced around $\phi = \pi$.
As shown in \cref{Fig7}, the probability density $\rho_n(y,z)$ confirms that the subgap states reside in the two superconducting blocks and are mostly localized along the $\hat y$ edges almost independently of the Josephson phase $\phi$.
However, the orientation of the local Majorana polarization vector, \cref{eq:localMP}, is strongly modified by $\phi$.
At $\phi=0$ ($\hat{\chi}_\nu\!\simeq\!1$) the local Majorana polarization vectors within each edge are
\emph{aligned}, so they add coherently [see arrows in panels \textbf{(b-g)}]. As the phase difference evolves towards $\phi\sim\pi$ the state localization remains mostly unchanged but the vectors at each edge \emph{rotate out of phase} and cancel, leading to $\hat{\chi}_\nu\to0$.
Consequently, the phase does not delocalize the state, it just turns the self-conjugate Majorana into a trivial Andreev state by disrupting the phase coherence of the local Majorana polarization.
In addition to the phase rotation of the local Majorana polarization vector, a careful inspection of the probability density shows that the subgap state with higher energy, labeled as $n=2$ in~\cref{Fig7}\textbf{(e-g)}, indeed delocalizes along the $\hat{y}$ direction edges when increasing $\phi$. Such delocalization is correlated with the phase-induced splitting away from zero-energy of this state; see \cref{Fig7}\textbf{(a)}.

In conclusion,
we expect the vertical modes of the quantum well to be reconstructed due to the Josephson phase in three main mechanisms:
A precession of the Majorana polarization in each upper and lower sectors leading to a total cancellation still preserving almost zero-energy bands [\cref{Fig6}\textbf{(c)}];
a gap opening due to the coupling of the vertical modes that form the \glspl{mzm}~[\cref{Fig6}\textbf{(e)}];
and a coupling of pairs of \glspl{mzm}~[\cref{Fig6}\textbf{(d, f)}].

The behavior with the phase difference is thus comparable to the effect of a linear \(k_z\) term discussed in \cref{sec:LinearZ}, but with important differences.
The linear term $B$ mixes orbital-parity sectors directly in the normal-state Hamiltonian and progressively destroys the quantum-well structure, whereas the Josephson phase $ \phi$ acts  through the superconducting pairing and reorganizes the existing Majorana channels into coupled doublets. The important consequence is that the phase difference can reversibly tune the effective Majorana-pair multiplicity without changing the slab thickness or the magnetic configuration.

\section{Conclusions}\label{sec:conclusions}
In this work we studied three-dimensional magnetic topological insulators proximitized by a conventional superconductor to create a platform for a multiple chiral Majorana phase. The multiple chiral modes are obtained through confinement effects in the vertical direction in a finite slab. In the limiting case of quantum-well quantization in the vertical direction, we found that it was possible to calculate the Chern number phase analytically using a simple argument that accounts for the parametric appearance of inverted subbands for permitted quantized momenta. We proved that this expression agreed with exact numerical diagonalization in finite systems.

We discussed more realistic scenarios that include usual Dirac linear coupling in the vertical stacking direction, as well as the effect of uncorrelated local disorder. In both cases, we demonstrated that the multiple chiral Majorana states remained unaffected by these additional terms over a wide range of parameters. Notably, the chiral topological states are resilient to Anderson disorder in the chemical potential landscape with an amplitude of $20\Delta$, demonstrating the robust topological protection of these states.
Finally, we considered non-uniform superconducting coupling in the vertical direction, examining a Josephson junction geometry with a relative phase between the two regions of the slab. This phase enables the number of chiral Majorana modes to be adjusted externally by hybridizing pairs of chiral Majorana channels, or by modifying the relative orientation of the local Majorana polarization.

\par The results presented here could provide a basis for designing multiterminal devices with tunable superconducting chiral channels that exploit alternating magnetically doped layers. From an experimental perspective,
multilayer topological-insulator heterostructures have already been realized without superconductors,
from thin slabs~\cite{Chang2013} to larger devices~\cite{Zhao2023}, while magnetic \cite{Zhuo2024} and electric fields~\cite{Baba2022, Yuan2023} provide additional tunability.
Beyond magnetically doped \glspl{ti},
the antiferromagnetic MnBi$_2$Te$_4$ offers another promising platform for related physics based on a recently observed strong even-odd layer effect arising from its layered antiferromagnetic order~\cite{Ovchinnikov2021, Mei2024, Yuan2024, Vyazovskaya2025}.
In both platforms, proximity-induced superconductivity has already shown encouraging initial results~\cite{Yi2024, Balakrishnan2025, Yuan2024}, pointing toward realistic routes for realizing the phases proposed here.

More broadly, our results reveal that although chiral Majorana states are localized near the sample surface, their structure across the device thickness governs their number, hybridization, and controllability. This additional degree of freedom opens new opportunities for exploiting the layered character of topological materials in superconducting architectures, particularly in multiterminal geometries where the vertical distribution of chiral modes may enable novel forms of nonlocal transport in topologically protected channels.

\section*{Data and code availability}
The data that support the findings of this study are openly available at the following URL:
\url{https://github.com/AlejandroSGomez/multiple-chiral-majorana-states-in-proximitized-magnetic-topological-insulator-heterostructures}.

\section*{Funding}
This work has been supported by
Spanish CM ``Talento Program'' project No.~2019-T1/IND-14088 and No.~2023-5A/IND-28927, the Agencia Estatal de Investigaci\'on (MCIN/AEI/10.13039/ 501100011033) project No.~PID2020-117992GA-I00, No.~CNS2022-135950, No.~PID2022-136285NB-C31, and No.~PID2024-157821NB-I00, and through the ``María de Maeztu'' Programme for Units of Excellence in R\&D (CEX2023-001316-M).

\appendix

\section{Analytical derivation of the two-dimensional limit}
\label{app:2d-derivation}

In this appendix we derive the gap-closing conditions presented in \cref{sec:2Dlimit}, \cref{eq:GapClosings}, based on the analytical solution of the secular equation of the eight component \gls{bdg} matrix, \cref{eq:FullBdG}.
For simplicity, we limit our analysis to the strictly two-dimensional model setting \(C_1=M_1=B=0\) in \cref{eq:FullBdG}.
Allowing for a finite \(M_1\) does not modify the analysis or the resulting gap-closing conditions. By contrast, a finite linear term \(B\) leads to a different behavior, which is discussed separately in \cref{sec:LinearZ}.
First, we introduce some useful identities for the Clifford algebra related to the $\Gamma_{ijk}$ matrices and a short-hand notation for the Hamiltonian, \cref{eq:FullBdG}. Next, we derive the analytical secular equation for the eigenvalue problem of \cref{eq:FullBdG} and, finally, evaluate its zero-energy solutions.

\subsection{Minimal Clifford-algebra identities}

The product of Pauli matrices, including the identity \(\sigma_0=\mathbb I_2\), is given by
\begin{equation}
    \sigma_i\sigma_j
    =
    \delta_{ij}\sigma_0
    +\mathrm{i}\epsilon_{ijk}\sigma_k,
    \qquad i,j\in\{x,y,z\}.
    \label{eq:app-pauli-product}
\end{equation}
Consequently, products of the matrices \(\hGamma{ijk}=\hat\tau_i\otimes\hat s_j\otimes\hat\sigma_k\), evaluated
component by component, result in
\begin{equation}
    \hGamma{ijk}\hGamma{lmn}
    =
    (\hat\tau_i\hat\tau_l)
    \otimes(\hat s_j\hat s_m)
    \otimes(\hat\sigma_k\hat\sigma_n).
    \label{eq:app-gamma-product}
\end{equation}
Then, any Clifford set \(\{A_i\}\) such that \(\{A_i,A_j\}=2\delta_{ij}\mathbb I\) fulfills the identity
\begin{equation}
    \Big(\sum_i a_iA_i\Big)^2
    =
    \sum_i a_i^2\mathbb I,
    \label{eq:app-clifford-square}
\end{equation}
with $a_j$ complex numbers, because every cross term is proportional to an anticommutator between distinct generators.

Additionally, if \(L\ket{\psi}=R\ket{\psi}\) then
\begin{equation}
    L^2\ket{\psi}
    =LR\ket{\psi}
    =\left(RL+[L,R]\right)\ket{\psi}
    =\left(R^2+[L,R]\right)\ket{\psi}.
    \label{eq:app-operator-step}
\end{equation}

We next apply identities \cref{eq:app-clifford-square,eq:app-operator-step} repeatedly to obtain the secular equation.

\subsection{Two-dimensional BdG Hamiltonian}

For ease of notation, we introduce the momentum-dependent abbreviations
\begin{align}
    \mathcal S_x&=\frac{A}{a}\sin(k_xa),
    &\mathcal S_y&=\frac{A}{a}\sin(k_ya),
    &\mathcal A_{\mathbf k}^2&=\mathcal S_x^2+\mathcal S_y^2.
    \label{eq:app-momentum-shorthand}
\end{align}
The two-dimensional Hamiltonian obtained from \cref{eq:FullBdG} is then written as
\begin{equation}
    \mathcal H_{\mathrm{2D}}(\mathbf k)
    =
    \varepsilon_{\mathbf k}\hGamma{z00}
    +\mathcal M_{\mathbf k}\hGamma{z0z}
    +\mathcal S_x\hGamma{00x}
    +\mathcal S_y\hGamma{zzy}
    +G\hGamma{zzz}
    +T\hGamma{zyy}
    +\Delta\hGamma{yyz}.
\label{eq:app-2d-bdg}
\end{equation}

\subsection{First squaring}

The eigenvalue equation \(\mathcal H_{\mathrm{2D}}\ket{\psi}=E\ket{\psi}\) can
be rearranged as
\begin{equation}
    L\ket{\psi}=R\ket{\psi},
    \label{eq:app-LR-eigenproblem}
\end{equation}
where
\begin{align}
    L&=E\hGamma{000}
       -\varepsilon_{\mathbf k}\hGamma{z00}
       -\Delta\hGamma{yyz},
       \label{eq:app-L-definition}\\
    R&=\mathcal M_{\mathbf k}\hGamma{z0z}
       +\mathcal S_x\hGamma{00x}
       +\mathcal S_y\hGamma{zzy}
       +G\hGamma{zzz}
       +T\hGamma{zyy}.
       \label{eq:app-R-definition}
\end{align}
Squaring the three terms in \(L\) results in
\begin{equation}
    L^2
    =
    (E^2+\varepsilon_{\mathbf k}^2+\Delta^2)\hGamma{000}
    -2E\left(
        \varepsilon\hGamma{z00}
        +\Delta\hGamma{yyz}
    \right),
    \label{eq:app-L-square}
\end{equation}
where the \(\varepsilon_{\mathbf k}\Delta\) cross term vanishes because
\(\{\hGamma{z00},\hGamma{yyz}\}=0\). Similarly expanding \(R^2\) reads
\begin{equation}
    R^2=
    (\mathcal M_{\mathbf k}^2+\mathcal S_x^2+\mathcal S_y^2+G^2+T^2)
    \hGamma{000}
    +2G\left(
        \mathcal M_{\mathbf k}\hGamma{0z0}
        -T\hGamma{0xx}
    \right).
\label{eq:app-R-square}
\end{equation}
All other cross terms in \(R^2\) vanish. For example,
\(\hGamma{z0z}\) and \(\hGamma{zzz}\) commute and multiply to
\(\hGamma{0z0}\), whereas \(\hGamma{zyy}\hGamma{zzz}=-\hGamma{0xx}\).

Finally, the only nonzero part of the commutator in \cref{eq:app-operator-step} comes from the pairing term,
\begin{equation}
    [L,R]
    =-2\mathrm{i}\Delta\left(
        \mathcal M_{\mathbf k}\hGamma{xy0}
        +\mathcal S_x\hGamma{yyy}
        +\mathcal S_y\hGamma{xxx}
    \right).
    \label{eq:app-LR-commutator}
\end{equation}

Substitution of \cref{eq:app-L-square,eq:app-R-square,eq:app-LR-commutator}
into \cref{eq:app-operator-step}, followed by moving all scalar terms to the
left-hand side, yields
\begin{equation}
\begin{aligned}
    \alpha(E)\hGamma{000}\ket{\psi}
    ={}&\Bigl[
    2E\left(
        \varepsilon_{\mathbf k}\hGamma{z00}
        +\Delta\hGamma{yyz}
    \right)
    +2G\left(
        \mathcal M_{\mathbf k}\hGamma{0z0}
        -T\hGamma{0xx}
    \right)
    \\
    &\quad
    -2\mathrm{i}\Delta\left(
        \mathcal M_{\mathbf k}\hGamma{xy0}
        +\mathcal S_x\hGamma{yyy}
        +\mathcal S_y\hGamma{xxx}
    \right)
    \Bigr]\ket{\psi},
\end{aligned}
\label{eq:app-after-first-square}
\end{equation}
with
\begin{equation}
    \alpha(E)
    =E^2+\varepsilon_{\mathbf k}^2+\Delta^2
    -\mathcal M_{\mathbf k}^2-\mathcal S_x^2-\mathcal S_y^2-G^2-T^2.
    \label{eq:app-alpha}
\end{equation}

\subsection{Second squaring}

\Cref{eq:app-after-first-square} is more compactly written as
\begin{equation}\label{eq:secondsquare}
    \alpha(E)\hGamma{000}\ket{\psi}=(2EP+2GQ-2\mathrm{i}\Delta C)\ket{\psi} ,
\end{equation}
with
\begin{equation}
    P=\varepsilon_{\mathbf k}\hGamma{z00}+\Delta\hGamma{yyz}, \quad
    Q=\mathcal M_{\mathbf k}\hGamma{0z0}-T\hGamma{0xx}, \quad
    C=\mathcal M_{\mathbf k}\hGamma{xy0}
       +\mathcal S_x\hGamma{yyy}
       +\mathcal S_y\hGamma{xxx}.
    \label{eq:app-PQC}
\end{equation}
Expressing the square of \cref{eq:secondsquare} again in the form of \cref{eq:app-operator-step} results in the terms
\begin{equation}
    P^2=(\varepsilon_{\mathbf k}^2+\Delta^2)\hGamma{000},\quad
    Q^2=(\mathcal M_{\mathbf k}^2+T^2)\hGamma{000},\quad
    C^2=(\mathcal M_{\mathbf k}^2+\mathcal S_x^2+\mathcal S_y^2)\hGamma{000},
    \label{eq:app-PQC-algebra}
\end{equation}
and mixed anticommutators
\begin{subequations}\label{eq:app-PQC-algebra2}
\begin{align}
    \{P,Q\}={}&2\left(
        \varepsilon_{\mathbf k}\mathcal M_{\mathbf k}\hGamma{zz0}
        -\varepsilon_{\mathbf k} T\hGamma{zxx}
        -T\Delta\hGamma{yzy}
    \right),\\
    \{P,C\}={}&0,\\
    \{Q,C\}={}&2T\left(
        \mathcal S_x\hGamma{yzz}
        -\mathcal S_y\hGamma{x00}
    \right).
\end{align}
\end{subequations}

As a result, the square of \cref{eq:secondsquare}, after inserting \cref{eq:app-PQC-algebra,eq:app-PQC-algebra2}, reads as
\begin{equation}
\begin{aligned}
    \alpha(E)^2\hGamma{000}\ket{\psi}
    ={}&\Bigl[
       4E^2(\varepsilon_{\mathbf k}^2+\Delta^2)
       +4G^2(\mathcal M_{\mathbf k}^2+T^2)
       -4\Delta^2(\mathcal M_{\mathbf k}^2+\mathcal S_x^2+\mathcal S_y^2)
    \Bigr]\hGamma{000}\ket{\psi}
    \\
    &+8EG\left(
        \varepsilon_{\mathbf k}\mathcal M_{\mathbf k}\hGamma{zz0}
        -\varepsilon_{\mathbf k} T\hGamma{zxx}
        -T\Delta\hGamma{yzy}
    \right)\ket{\psi}
    \\
    &-8\mathrm{i}GT\Delta\left(
        \mathcal S_x\hGamma{yzz}
        -\mathcal S_y\hGamma{x00}
    \right)\ket{\psi}.
\end{aligned}
\label{eq:app-second-square}
\end{equation}
We identify the scalar part as
\begin{equation}
    \beta(E)=
       4G^2(\mathcal M_{\mathbf k}^2+T^2)
       +4E^2(\varepsilon_{\mathbf k}^2+\Delta^2)
       -4\Delta^2(\mathcal M_{\mathbf k}^2+\mathcal S_x^2+\mathcal S_y^2).
\label{eq:app-beta}
\end{equation}
Moving this scalar term to the left-hand side results in
\begin{equation}
\begin{aligned}
    \left[\alpha(E)^2-\beta(E) \right]\hGamma{000}\ket{\psi} =
    {}&
    8EG\left(
        \varepsilon_{\mathbf k}\mathcal M_{\mathbf k}\hGamma{zz0}
        -\varepsilon_{\mathbf k} T\hGamma{zxx}
        -T\Delta\hGamma{yzy}
    \right)
    \ket{\psi}
    \\
    &
    -8\mathrm{i}GT\Delta\left(
        \mathcal S_x\hGamma{yzz}
        -\mathcal S_y\hGamma{x00}
    \right)
    \ket{\psi}.
\end{aligned}
\label{eq:app-regrouped}
\end{equation}

\begin{figure}
    \centering
    \includegraphics[width=\linewidth]{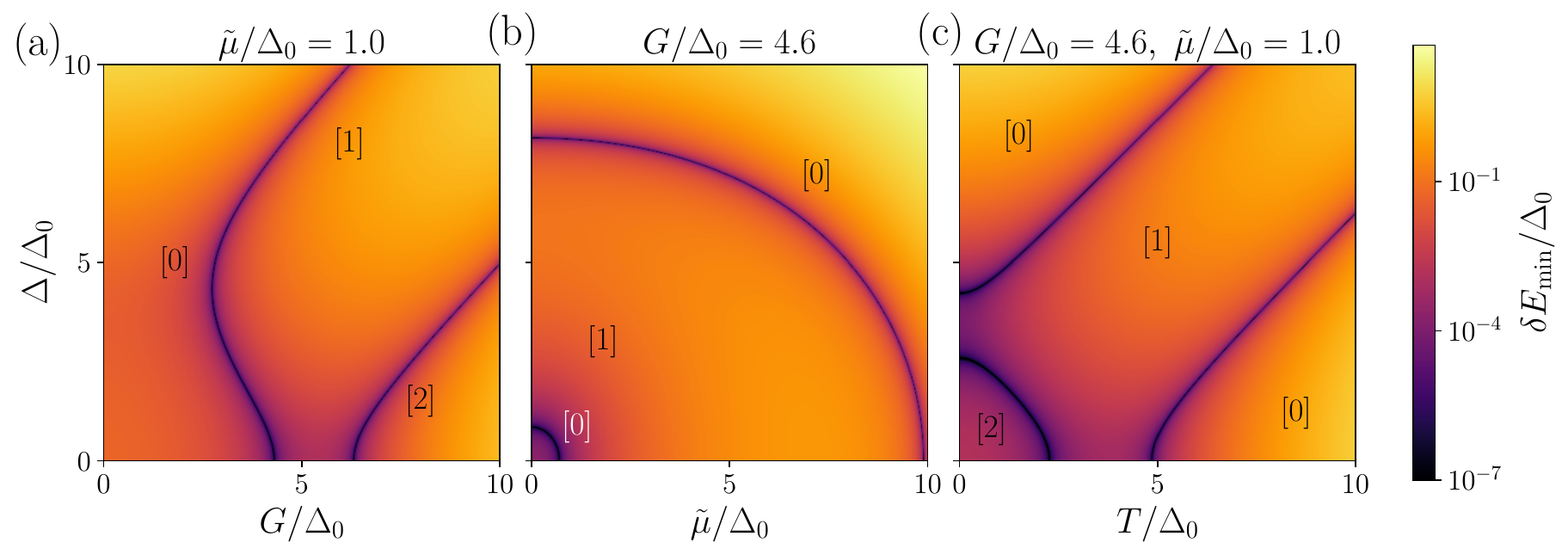}
    \caption{
    \textbf{Parameter dependence of the two-dimensional phase diagram.}
    Maps of the smallest absolute BdG eigenenergy $\delta E_{\min}/\Delta_0$, obtained by numerical diagonalization, as a function of the pairing amplitude $\Delta$ and
    \textbf{(a)} \(G\) fixed \(\tilde{\mu}/\Delta_0=1\) and \(T/\Delta_0=4.5\);
    \textbf{(b)} \(\tilde{\mu}\) at \(G/\Delta_0=4.6\) and \(T/\Delta_0=4.5\); and
    \textbf{(c)} \(T\) for \(G/\Delta_0=4.6\) and \(\tilde{\mu}/\Delta_0=1\).
    Black contours mark the analytical gap-closing boundaries of \cref{eq:GapClosings}, and \(|C|\) is indicated for each region in brackets.
    We defined $\tilde{\mu} \equiv \mu+C_0$ and normalized all energies to \(\Delta_0\), the value used for the pairing amplitude in the main text.
    }
    \label{Fig1Ap}
\end{figure}

\subsection{Final squaring and secular equation}

The five matrices
\begin{equation}
    \left\{
    \hGamma{zz0},\hGamma{zxx},\hGamma{yzy},
    \hGamma{yzz},\hGamma{x00}
    \right\}
    \label{eq:app-final-clifford-set}
\end{equation}
form a Clifford set. Squaring \cref{eq:app-regrouped} thus removes the remaining matrix structure.
We take this into account to obtain
\begin{equation}
    [\alpha(E)^2-\beta(E)]^2\hGamma{000}\ket{\psi}
    =\gamma(E)\hGamma{000}\ket{\psi},
    \label{eq:app-final-square}
\end{equation}
with
\begin{equation}
    \gamma(E)=
    64E^2G^2\left(
        \varepsilon_{\mathbf k}^2\mathcal M_{\mathbf k}^2
        +\varepsilon_{\mathbf k}^2T^2
        +T^2\Delta^2
    \right)
    -64G^2T^2\Delta^2(\mathcal S_x^2+\mathcal S_y^2).
\label{eq:app-gamma}
\end{equation}
The eight Bogoliubov bands are the roots of the scalar secular equation
\begin{equation}
    \boxed{
    \left[\alpha(E,\mathbf k)^2-\beta(E,\mathbf k)\right]^2
    =\gamma(E,\mathbf k)
    },
    \label{eq:app-secular-equation}
\end{equation}
where \(\alpha\), \(\beta\), and \(\gamma\) are given explicitly in
\cref{eq:app-alpha,eq:app-beta,eq:app-gamma}, respectively.
No matrix elements remain in
\cref{eq:app-secular-equation}; it is an eighth-order scalar polynomial in
\(E\), with the particle-hole-related roots appearing in \(\pm E\) pairs.

\subsection{Zero-energy gap-closing conditions}

Setting \(E=0\) in \cref{eq:app-secular-equation} results in
\begin{equation}
    [\alpha(0,\mathbf k)^2-\beta(0,\mathbf k)]^2
    =-64G^2T^2\Delta^2\mathcal A_{\mathbf k}^2,
    \label{eq:app-zero-energy-equation}
\end{equation}
with
\begin{equation}
    A_{\mathbf k}^2 = \mathcal S_x^2+\mathcal S_y^2 .
\end{equation}
The superconducting regime considered in the main text thus requires that \(G T\Delta\neq0\).
The left-hand side of
\cref{eq:app-zero-energy-equation} is nonnegative while its right-hand side is
nonpositive. Equality is, therefore, only possible when both sides vanish:
\begin{align}
    \mathcal A_{\mathbf k}^2&=0,
    \label{eq:app-gap-condition-1}\\
    \alpha(0,\mathbf k)^2&=\beta(0,\mathbf k).
    \label{eq:app-gap-condition-2}
\end{align}
The first condition reads
\begin{equation}
    \sin^2(k_xa)+\sin^2(k_ya)=0,
    \label{eq:app-trim-condition}
\end{equation}
and hence restricts a gap closing to a time-reversal-invariant momentum
\(\mathbf k=\Lambda_i\). For such momenta we get
\(\mathcal S_x=\mathcal S_y=0\), and the second condition becomes
\begin{equation}
\begin{aligned}
    \bigl[
    \varepsilon_{\Lambda_i}^2+\Delta^2
    -\mathcal M_{\Lambda_i}^2-G^2-T^2
    \bigr]^2
    ={}&4\mathcal M_{\Lambda_i}^2(G^2-\Delta^2)
    +4G^2T^2,
\end{aligned}
\label{eq:app-gap-closing}
\end{equation}
which is precisely \cref{eq:GapClosings} of the main text.

For completeness, the explicit boundary in \(G^2\) follows without any further
matrix algebra. Setting \(x=G^2\) and
\begin{equation}
    U=\varepsilon_{\Lambda_i}^2+\Delta^2
      -\mathcal M_{\Lambda_i}^2-T^2,
    \label{eq:app-U-definition}
\end{equation}
\Cref{eq:app-gap-closing} reads as
\begin{equation}
    x^2
    -2\left(
       \varepsilon_{\Lambda_i}^2+\Delta^2
       +\mathcal M_{\Lambda_i}^2+T^2
    \right)x
    +U^2+4\mathcal M_{\Lambda_i}^2\Delta^2=0 ,
    \label{eq:app-G-quadratic}
\end{equation}
with solutions
\begin{equation}
    G^2=
    \Delta^2+\varepsilon_{\Lambda_i}^2
    +\mathcal M_{\Lambda_i}^2+T^2
    \pm2\sqrt{
       \varepsilon_{\Lambda_i}^2
       (\mathcal M_{\Lambda_i}^2+T^2)
       +T^2\Delta^2
    }.
\label{eq:app-G-boundary}
\end{equation}
At the Gamma point \(\Lambda_i=\Gamma\), we have \(\varepsilon_\Gamma=C_0+\mu\equiv\tilde\mu\) and \cref{eq:app-G-boundary} reduces directly to \cref{eq:boundary-C0fixed-2}.
The resulting analytical phase boundaries are shown in \cref{Fig1Ap} for representative two-dimensional cuts of the \(\{G,\Delta,T,\tilde\mu\}\) parameter space; the dark contours coincide with the zero-energy conditions derived above.

Note that if either one of \(G\), \(T\), or \(\Delta\) vanishes, the derivation of \cref{eq:app-gap-condition-1} from \cref{eq:app-zero-energy-equation} must be reconsidered separately. The resulting singular limits are not the chiral superconducting regime analyzed here.

\section{Majorana pair counting at the particle-hole-invariant momentum \texorpdfstring{$k_x=0$}{kx=0}}
\label{app:linear-kx0}

\begin{figure}[htb]
    \centering
    \includegraphics[width=\linewidth]{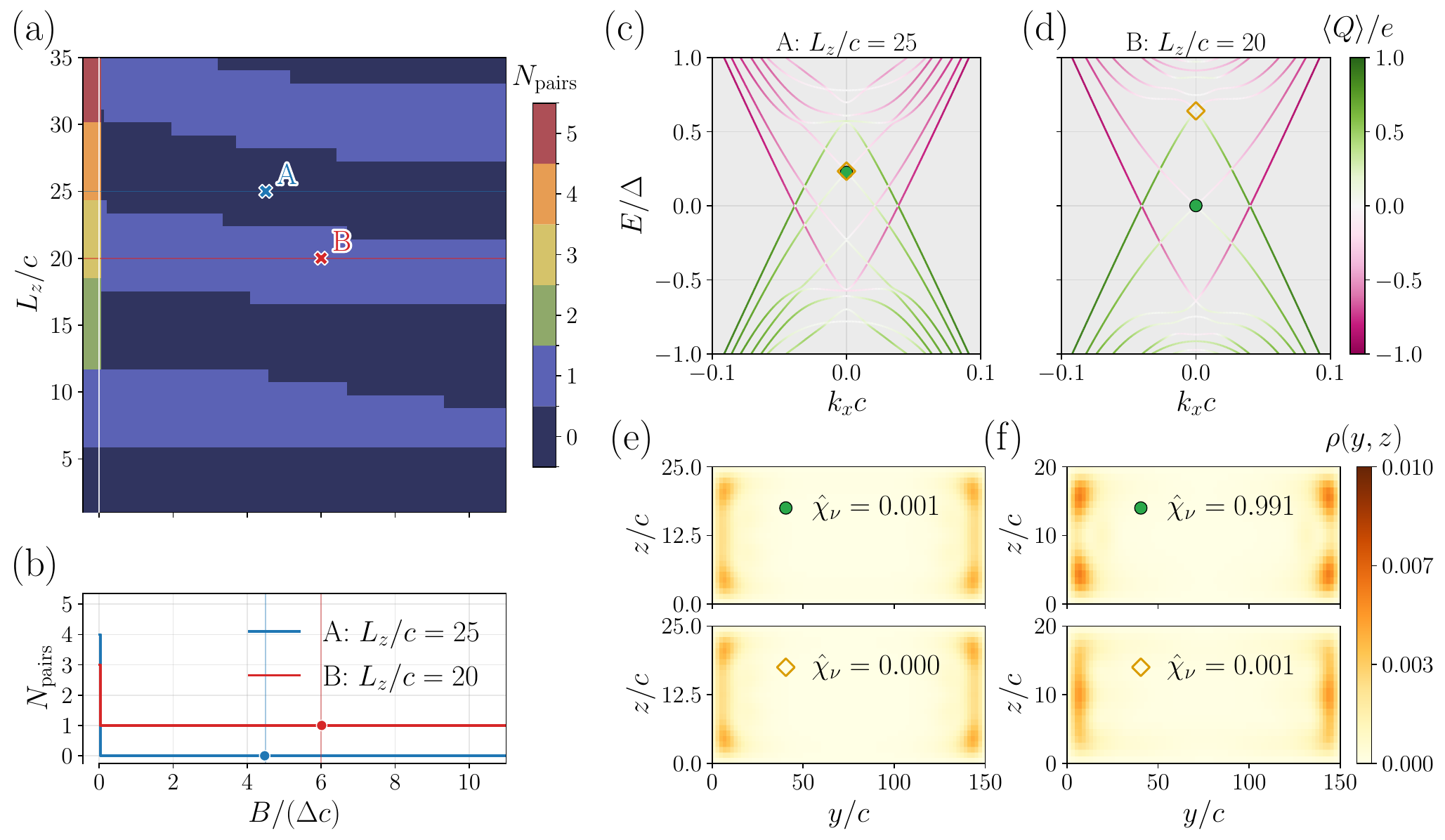}
    \caption{\textbf{Same quantities as \cref{Fig4} evaluated
    exactly at $k_x=0$ using the edge-resolved Majorana indicator.}
    All panels use the same slab model, parameter ranges, and selection rule as~\cref{Fig4}.
    \textbf{(a)} Map of $N_{\rm pairs}$ as a function of $B$ and $L_z$; the unlabelled strip to the left of the white boundary line repeats the $B=0$ column for visualization only. \textbf{(b)} Cuts at $L_z/c=25$ (A) and $L_z/c=20$ (B).
    \textbf{(c,d)} Low-energy spectra for the indicated points in \textbf{(a)},
    coloured by the normalized charge expectation \(\langle Q\rangle/e\).
    For $k_xc=0$ the lowest non-negative-energy states are marked by green circles and the next positive-energy bands by gold diamonds.
    The marked points have $(E/\Delta,\hat{\chi}_\nu)$ values equal to $(0.226,0.001)$ and $(0.234,0.000)$ at A, and $(0.0019,0.991)$ and $(0.640,0.001)$ at B.
    \textbf{(e,f)} Probability densities $\rho(y,z)$ for the two marked
    states, with the green state at the top and the gold one at the bottom, as indicated beside the corresponding $\hat{\chi}_\nu$ value.
    }
    \label{Fig2Ap}
\end{figure}

\Cref{Fig4} in the main text was computed for a finite, positive momentum $k_x c\simeq6.7\times10^{-3}$, instead of at exactly the particle-hole-invariant point $k_x=0$.
We now justify this choice by repeating in \cref{Fig2Ap} the exact same calculation for $k_x=0$ (same slab, same ranges $B/(\Delta c)\in[0,11]$ and $L_z/c\in[1,35]$, and the same thresholds).

A Majorana character indicator must be \emph{edge-resolved}, i.e., sensitive to the specific localization of the Majorana state at any momenta. However, while at $k_x=0$ the counter-propagating chiral modes bound to opposite $\hat y$ boundaries, they also span an exactly degenerate subspace. The eigenvectors within this subspace are not unique, and a numerical diagonalization may return arbitrary linear combinations of the two edge-localized states. In an edge-localized basis, the two boundaries contribute opposite Majorana polarizations; consequently, the sample-wide self-conjugacy amplitude $\sum_{\mathbf r}{\rm MP}_\nu(\mathbf r)$ can cancel and does not provide a basis-stable diagnostic of either edge.
Evaluating the polarization separately in $\Omega_L$ and $\Omega_R$ through $\chi^{L/R}_\nu$ [\cref{eq:regional-majorana-polarization}] avoids this inter-edge cancellation.

Using the same edge-resolved criteria of \cref{Fig4} of the main text, the $B=0$ limit reproduces the quantum-well staircase of~\cref{sec:Stacking} exactly [\cref{Fig2Ap}(a), leftmost column].
Indeed, $N_{\rm pairs}$ grows by one each time an additional confined subband crosses zero, from a single pair near $L_z/c\simeq6$ to five pairs at $L_z/c\simeq35$, in one-to-one correspondence with the $k_x c=0^{+}$ count and with $\hat{\chi}_\nu\simeq1$ for every counted state.

By contrast, at finite $B$ the count evaluated strictly at $k_x=0$ is much more sensitive to the phase structure of the eigenstates. The linear term enters the $k_x=0$ Bloch Hamiltonian through the purely imaginary hopping $\mathrm{i}\,[B/(2\Delta c)]\,\hGamma{0yy}$ [cf.~\cref{eq:FullBdG}] and changes the relative phases of the wavefunction components. The phase of the resulting local Majorana polarization can rotate along the boundary, suppressing the coherent edge sum $|\sum_{\mathbf r\in\Omega}{\rm MP}_\nu|$ even when a state remains edge-localized and close to zero energy. Consequently, the pair count rule at exactly $k_x=0$ is no longer sensitive to the presence of multiple Majorana modes, already for $B/(\Delta c)\gtrsim5\times10^{-2}$.
As a result, the high-Chern number regions are misclassified for $B\neq0$ as single-pair cases [\cref{Fig2Ap}\textbf{(a,b)}].
That the counting rule is ill-defined for $k_x=0$ should not be interpreted as the disappearance of the chiral Majorana branches or as a change of bulk topology; it just reveals the phase sensitivity of the polarization-coherence diagnostic at $k_x=0$.

The representative points A and B illustrate this issue.
At point B of \cref{Fig2Ap}\textbf{(a)}, the criterion identifies only one pair at $k_x=0$ and $B\neq0$. We mark in the corresponding band dispersion of \cref{Fig2Ap}\textbf{(d)} the lowest nonnegative-energy state with $E/\Delta\simeq2\times10^{-3}$ and $\hat{\chi}_\nu\simeq0.99$ with a green circle, whereas the next positive-energy band lies at $E/\Delta\simeq0.64$ with $\hat{\chi}_\nu\simeq10^{-3}$.
At point A of \cref{Fig2Ap}\textbf{(a)}, the two lowest positive-energy states lie at $E/\Delta\simeq0.226$ and $0.234$, both with $\hat{\chi}_\nu\simeq10^{-3}$ or smaller. Therefore, no pair is counted at that momentum [\cref{Fig2Ap}\textbf{(c,e)}]. Nevertheless, the spectra at both points exhibit an off-zero crossing, reinforcing the result that $N_{\rm pairs}$ in~\cref{Fig2Ap}\textbf{(a)} is a momentum-resolved classification rather than a bulk invariant.

While the case with $k_x=0$ is singular for the pair counting rule, for $k_x=0^{+}$ the opposite-edge degeneracy is lifted and the chiral branches can be resolved on individual boundaries. The pair count then remains stable over the wide range of $B$ shown in~\cref{Fig4}. Evaluating the count at the first positive momentum of the numerical grid thus avoids the degenerate-subspace ambiguity while remaining arbitrarily close to the particle-hole-invariant point. This is the convention adopted in the main text.

\section{\texorpdfstring{Josephson-phase control and spectral
periodicity}{Josephson-phase control and spectral periodicity}}
\label{app:4pi}

The vertical Josephson junction of Sec.~\ref{sec:JJ} uses the phase
difference as a control parameter for the hybridization and multiplicity of the
Majorana channels. Here we establish the spectral periodicity of this response
using the low-energy gap and its finite-size scaling, and connect it to the
number of Majorana-like pairs. The geometry, parameters, and phase profile
\(\Delta(z)\) are those of Sec.~\ref{sec:JJ}.

The phase of the complex Majorana polarization is not itself an observable.
Under the arbitrary rephasing of a \gls{bdg} eigenstate,
\(\psi_\nu\to e^{\mathrm{i}\theta}\psi_\nu\), the bilinear local polarization
transforms as
\begin{equation}
    \mathrm{MP}_\nu(\mathbf r)\longrightarrow
    e^{2\mathrm{i}\theta}\mathrm{MP}_\nu(\mathbf r).
    \label{eq:mp-gauge}
\end{equation}
Consequently, a sign or phase change of \(\mathrm{MP}_\nu\) after a
\(2\pi\) cycle is gauge dependent. The regional indicator
\(\chi_\nu(\Omega)\) (and hence the edge-resolved \(\hat{\chi}_\nu\)) is gauge
invariant and remains a valid measure of Majorana character. It does not,
however, determine the spectral periodicity.

The static Hamiltonian and its spectrum satisfy
\(H_{\mathrm{BdG}}(\phi+2\pi)=H_{\mathrm{BdG}}(\phi)\). A fractional Josephson
response would instead require robust spectral flow through zero energy, stable
against increasing the system size. The gap diagnostics below directly test
this requirement without assigning physical meaning to the gauge-dependent
phase of an individual eigenvector.

\begin{figure}[htb]
    \centering
    \includegraphics[width=\linewidth]{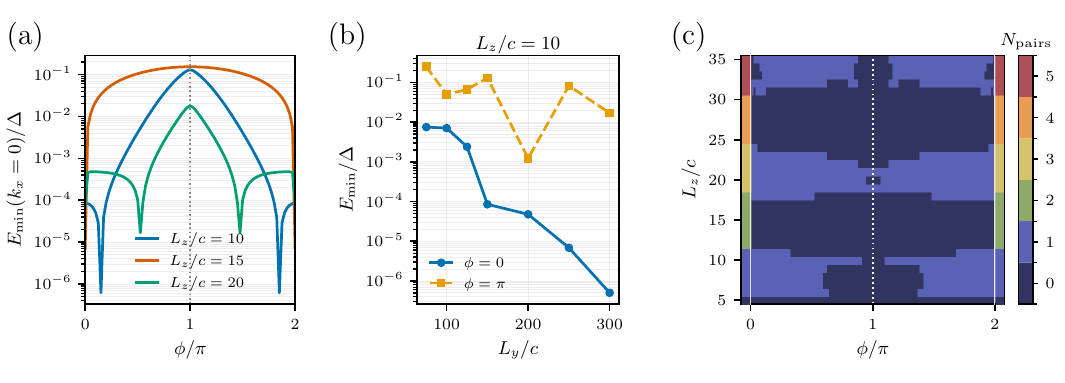}
    \caption{\textbf{Low-energy gap and Majorana-pair count of the vertical
    Josephson junction at \(k_x=0\).}
    \textbf{(a)} Minimum excitation energy \(E_{\min}/\Delta\) versus
    \(\phi/\pi\) for \(L_z/c=10,15,20\) at \(L_y/c=150\).
    \textbf{(b)} Finite-size evolution for \(L_z/c=10\): the residual splitting
    at \(\phi=0\) decreases with \(L_y\), whereas the \(\phi=\pi\) gap does not
    close systematically.
    \textbf{(c)} Thresholded pair count \(N_{\mathrm{pairs}}\) over the
    \((L_z/c,\phi)\) plane at \(L_y/c=150\), shown with a discrete colour scale.
    The unlabelled strips outside the white boundary lines repeat the \(\phi=0\) and \(2\pi\) columns for visualization only.
    The count is one half of the number of the fourteen states closest to zero energy that
    satisfy \(\hat{\chi}_\nu>0.5\) and
    \(|\langle Q_\nu\rangle|<0.1e\), as in \cref{Fig2Ap}.
    The dotted line marks \(\phi=\pi\).}
    \label{Fig3Ap}
\end{figure}

\Cref{Fig3Ap} summarizes the result. For
\(L_z/c=10,15,20\), the minimum \(k_x=0\) excitation gap remains finite and is
generally largest near \(\phi=\pi\) [panel \textbf{(a)}]. The narrow near-zero minima away
from \(\phi=\pi\) originate from the residual overlap between opposite-edge modes.
For \(L_z/c=10\), their splitting at \(\phi=0\) decreases approximately
exponentially with \(L_y\), whereas the gap at \(\phi=\pi\) shows no systematic
closure and remains much larger at the longest sizes [panel \textbf{(b)}]. The evolution
therefore does not exhibit a size-stable zero-energy crossing at \(\phi=\pi\).

The discrete map in \cref{Fig3Ap}\textbf{(c)} shows the complementary
thresholded pair count at \(k_x=0\). At \(\phi=0\), increasing \(L_z\) recovers
the quantum-well staircase, with one additional pair for each confined subband.
Changing the phase hybridizes these channels and reduces \(N_{\mathrm{pairs}}\)
toward \(\phi=\pi\) over a thickness-dependent window. Thus the Josephson phase
remains an effective control knob for the low-energy Majorana multiplicity even
though the spectrum does not display a protected \(4\pi\) response.

\begin{thebibliography}{10}

\bibitem{Hasan2010}
M.~Z. Hasan and C.~L. Kane.
\newblock Colloquium: Topological insulators.
\newblock {\em Rev. Mod. Phys.}, 82:3045--3067, Nov 2010.
\newblock \href {https://doi.org/10.1103/RevModPhys.82.3045}
  {\path{doi:10.1103/RevModPhys.82.3045}}.

\bibitem{Qi2011}
Xiao-Liang Qi and Shou-Cheng Zhang.
\newblock Topological insulators and superconductors.
\newblock {\em Reviews of Modern Physics}, 83(4):1057--1110, 2011.
\newblock \href {https://doi.org/10.1103/RevModPhys.83.1057}
  {\path{doi:10.1103/RevModPhys.83.1057}}.

\bibitem{Chang2013}
Cui-Zu Chang, Jinsong Zhang, Xiao Feng, Jie Shen, Zuocheng Zhang, Minghua Guo,
  Kang Li, Yunbo Ou, Pang Wei, Li-Li Wang, Zhong-Qing Ji, Yang Feng, Shuaihua
  Ji, Xi~Chen, Jinfeng Jia, Xi~Dai, Zhong Fang, Shou-Cheng Zhang, Ke~He, Yayu
  Wang, Li~Lu, Xu-Cun Ma, and Qi-Kun Xue.
\newblock Experimental observation of the quantum anomalous hall effect in a
  magnetic topological insulator.
\newblock {\em Science}, 340(6129):167--170, 2013.
\newblock \href {https://doi.org/10.1126/science.1234414}
  {\path{doi:10.1126/science.1234414}}.

\bibitem{Yuan2023}
Wei Yuan, Ling-Jie Zhou, Kaijie Yang, Yi-Fan Zhao, Ruoxi Zhang, Zijie Yan, Deyi
  Zhuo, Ruobing Mei, Yang Wang, Hemian Yi, Moses H.~W. Chan, Morteza Kayyalha,
  Chao-Xing Liu, and Cui-Zu Chang.
\newblock Electrical switching of the edge current chirality in quantum
  anomalous hall insulators.
\newblock {\em Nature Materials}, 23(1):58--64, October 2023.
\newblock \href {https://doi.org/10.1038/s41563-023-01694-y}
  {\path{doi:10.1038/s41563-023-01694-y}}.

\bibitem{Zhuo2024}
Deyi Zhuo, Lingjie Zhou, Yi-Fan Zhao, Ruoxi Zhang, Zi-Jie Yan, Annie~G. Wang,
  Moses H.~W. Chan, Chao-Xing Liu, Chui-Zhen Chen, and Cui-Zu Chang.
\newblock Engineering plateau phase transition in quantum anomalous hall
  multilayers.
\newblock {\em Nano Letters}, 24(23):6974--6980, 2024.
\newblock \href {https://doi.org/10.1021/acs.nanolett.4c01313}
  {\path{doi:10.1021/acs.nanolett.4c01313}}.

\bibitem{Kayyalha2020}
Morteza Kayyalha, Di~Xiao, Ruoxi Zhang, Jaeho Shin, Jue Jiang, Fei Wang, Yi-Fan
  Zhao, Run Xiao, Ling Zhang, Kajetan~M. Fijalkowski, Pankaj Mandal, Martin
  Winnerlein, Charles Gould, Qi~Li, Laurens~W. Molenkamp, Moses H.~W. Chan,
  Nitin Samarth, and Cui-Zu Chang.
\newblock Absence of evidence for chiral majorana modes in quantum anomalous
  hall-superconductor devices.
\newblock {\em Science}, 367(6473):64--67, January 2020.
\newblock \href {https://doi.org/10.1126/science.aax6361}
  {\path{doi:10.1126/science.aax6361}}.

\bibitem{Uday2024}
Anjana Uday, Gertjan Lippertz, Kristof Moors, Henry~F. Legg, Rikkie Joris,
  Andrea Bliesener, Lino M.~C. Pereira, A.~A. Taskin, and Yoichi Ando.
\newblock Induced superconducting correlations in a quantum anomalous hall
  insulator.
\newblock {\em Nature Physics}, pages 1--7, 2024.
\newblock \href {https://doi.org/10.1038/s41567-024-02574-1}
  {\path{doi:10.1038/s41567-024-02574-1}}.

\bibitem{Yi2024}
Hemian Yi, Yi-Fan Zhao, Ying-Ting Chan, Jiaqi Cai, Ruobing Mei, Xianxin Wu,
  Zi-Jie Yan, Ling-Jie Zhou, Ruoxi Zhang, Zihao Wang, Stephen Paolini, Run
  Xiao, Ke~Wang, Anthony~R. Richardella, John Singleton, Laurel~E. Winter,
  Thomas Prokscha, Zaher Salman, Andreas Suter, Purnima~P. Balakrishnan,
  Alexander~J. Grutter, Moses H.~W. Chan, Nitin Samarth, Xiaodong Xu, Weida Wu,
  Chao-Xing Liu, and Cui-Zu Chang.
\newblock Interface-induced superconductivity in magnetic topological
  insulators.
\newblock {\em Science}, 383(6683):634--639, February 2024.
\newblock \href {https://doi.org/10.1126/science.adk1270}
  {\path{doi:10.1126/science.adk1270}}.

\bibitem{Balakrishnan2025}
Purnima~P. Balakrishnan, Hemian Yi, Zi-Jie Yan, Wei Yuan, Andreas Suter,
  Christopher~J. Jensen, Pascal Manuel, Fabio Orlandi, Takayasu Hanashima,
  Christy~J. Kinane, Andrew~J. Caruana, Dirk Backes, Padraic Shafer, Brian~B.
  Maranville, Zaher Salman, Thomas Prokscha, Cui-Zu Chang, and Alexander~J.
  Grutter.
\newblock Depth-resolved magnetic order in superconducting topological
  insulator/fete thin film heterostructures.
\newblock {\em Physical Review Materials}, 9(10), October 2025.
\newblock \href {https://doi.org/10.1103/15fx-3cr2}
  {\path{doi:10.1103/15fx-3cr2}}.

\bibitem{Kitaev2001}
A.~Y. Kitaev.
\newblock Physics-uspekhi.
\newblock {\em 44}, 131., 2001.

\bibitem{Nayak2008}
Chetan Nayak, Steven~H. Simon, Ady Stern, Michael Freedman, and Sankar
  Das~Sarma.
\newblock Non-abelian anyons and topological quantum computation.
\newblock {\em Rev. Mod. Phys.}, 80:1083--1159, Sep 2008.
\newblock \href {https://doi.org/10.1103/RevModPhys.80.1083}
  {\path{doi:10.1103/RevModPhys.80.1083}}.

\bibitem{Alicea2012}
Jason Alicea.
\newblock New directions in the pursuit of majorana fermions in solid state
  systems.
\newblock {\em Reports on Progress in Physics}, 75(7):076501, jun 2012.
\newblock \href {https://doi.org/10.1088/0034-4885/75/7/076501}
  {\path{doi:10.1088/0034-4885/75/7/076501}}.

\bibitem{Beenakker2013}
C.W.J. Beenakker.
\newblock Search for majorana fermions in superconductors.
\newblock {\em Annual Review of Condensed Matter Physics}, 4(1):113--136, April
  2013.
\newblock \href {https://doi.org/10.1146/annurev-conmatphys-030212-184337}
  {\path{doi:10.1146/annurev-conmatphys-030212-184337}}.

\bibitem{Sarma2015}
Sankar~Das Sarma, Michael Freedman, and Chetan Nayak.
\newblock Majorana zero modes and topological quantum computation.
\newblock {\em npj Quantum Information}, 1(1):15001, Oct 2015.
\newblock \href {https://doi.org/10.1038/npjqi.2015.1}
  {\path{doi:10.1038/npjqi.2015.1}}.

\bibitem{Kouwenhoven2025}
Leo Kouwenhoven.
\newblock Perspective on majorana bound-states in hybrid
  superconductor-semiconductor nanowires.
\newblock {\em Modern Physics Letters B}, 39(03):2540002, 2025.
\newblock \href {https://doi.org/10.1142/S0217984925400020}
  {\path{doi:10.1142/S0217984925400020}}.

\bibitem{Qi2010}
Xiao-Liang Qi, Taylor~L. Hughes, and Shou-Cheng Zhang.
\newblock Chiral topological superconductor from the quantum hall state.
\newblock {\em Physical Review B}, 82(18):184516, November 2010.
\newblock \href {https://doi.org/10.1103/physrevb.82.184516}
  {\path{doi:10.1103/physrevb.82.184516}}.

\bibitem{Fu2008}
Liang Fu and Charles~L Kane.
\newblock Superconducting proximity effect and majorana fermions at the surface
  of a topological insulator.
\newblock {\em Physical Review Letters}, 100(9):096407, 2008.
\newblock \href {https://doi.org/10.1103/PhysRevLett.100.096407}
  {\path{doi:10.1103/PhysRevLett.100.096407}}.

\bibitem{Lutchyn2010}
Roman~M. Lutchyn, Jay~D. Sau, and S.~Das~Sarma.
\newblock Majorana fermions and a topological phase transition in
  semiconductor-superconductor heterostructures.
\newblock {\em Phys. Rev. Lett.}, 105:077001, Aug 2010.
\newblock \href {https://doi.org/10.1103/PhysRevLett.105.077001}
  {\path{doi:10.1103/PhysRevLett.105.077001}}.

\bibitem{Oreg2010}
Yuval Oreg, Gil Refael, and Felix von Oppen.
\newblock Helical liquids and majorana bound states in quantum wires.
\newblock {\em Phys. Rev. Lett.}, 105:177002, Oct 2010.
\newblock \href {https://doi.org/10.1103/PhysRevLett.105.177002}
  {\path{doi:10.1103/PhysRevLett.105.177002}}.

\bibitem{He2014}
James~J. He, Jiansheng Wu, Ting-Pong Choy, Xiong-Jun Liu, Y.~Tanaka, and K.~T.
  Law.
\newblock Correlated spin currents generated by resonant-crossed andreev
  reflections in topological superconductors.
\newblock {\em Nature Communications}, 5(1), February 2014.
\newblock \href {https://doi.org/10.1038/ncomms4232}
  {\path{doi:10.1038/ncomms4232}}.

\bibitem{DiMiceli2023}
Daniele Di~Miceli, Eduárd Zsurka, Julian Legendre, Kristof Moors, Thomas~L.
  Schmidt, and Llorenç Serra.
\newblock Conductance asymmetry in proximitized magnetic topological insulator
  junctions with majorana modes.
\newblock {\em Physical Review B}, 108(3):035424, July 2023.
\newblock \href {https://doi.org/10.1103/physrevb.108.035424}
  {\path{doi:10.1103/physrevb.108.035424}}.

\bibitem{Wang2015}
Jing Wang, Quan Zhou, Biao Lian, and Shou-Cheng Zhang.
\newblock Chiral topological superconductor and half-integer conductance
  plateau from quantum anomalous hall plateau transition.
\newblock {\em Physical Review B}, 92(6):064520, August 2015.
\newblock \href {https://doi.org/10.1103/physrevb.92.064520}
  {\path{doi:10.1103/physrevb.92.064520}}.

\bibitem{Lian2018}
Biao Lian, Jing Wang, Xiao-Qi Sun, Abolhassan Vaezi, and Shou-Cheng Zhang.
\newblock Quantum phase transition of chiral majorana fermions in the presence
  of disorder.
\newblock {\em Physical Review B}, 97(12):125408, March 2018.
\newblock \href {https://doi.org/10.1103/physrevb.97.125408}
  {\path{doi:10.1103/physrevb.97.125408}}.

\bibitem{Huang2018}
Yingyi Huang, F.~Setiawan, and Jay~D. Sau.
\newblock Disorder-induced half-integer quantized conductance plateau in
  quantum anomalous hall insulator-superconductor structures.
\newblock {\em Physical Review B}, 97(10):100501, March 2018.
\newblock \href {https://doi.org/10.1103/physrevb.97.100501}
  {\path{doi:10.1103/physrevb.97.100501}}.

\bibitem{Zhang2020}
Jian-Xiao Zhang and Chao-Xing Liu.
\newblock Disordered quantum transport in quantum anomalous hall
  insulator-superconductor junctions.
\newblock {\em Physical Review B}, 102(14):144513, October 2020.
\newblock \href {https://doi.org/10.1103/physrevb.102.144513}
  {\path{doi:10.1103/physrevb.102.144513}}.

\bibitem{Wang2018}
Jing Wang and Biao Lian.
\newblock Multiple chiral majorana fermion modes and quantum transport.
\newblock {\em Physical Review Letters}, 121(25):256801, December 2018.
\newblock \href {https://doi.org/10.1103/physrevlett.121.256801}
  {\path{doi:10.1103/physrevlett.121.256801}}.

\bibitem{Legendre2024}
Julian Legendre, Eduárd Zsurka, Daniele Di~Miceli, Llorenç Serra, Kristof
  Moors, and Thomas~L. Schmidt.
\newblock Topological properties of finite-size heterostructures of magnetic
  topological insulators and superconductors.
\newblock {\em Physical Review B}, 110(7):075426, 2024.
\newblock \href {https://doi.org/10.1103/PhysRevB.110.075426}
  {\path{doi:10.1103/PhysRevB.110.075426}}.

\bibitem{Zsurka2025}
Eduárd Zsurka, Daniele Di~Miceli, Julian Legendre, Llorenc Serra, Detlev
  Grützmacher, Thomas~L. Schmidt, and Kristof Moors.
\newblock Optimizing proximitized magnetic topological insulator nanoribbons
  for majorana bound states, may 2025.
\newblock \href {https://arxiv.org/abs/2505.02163} {\path{arXiv:2505.02163}},
  \href {https://doi.org/10.48550/ARXIV.2505.02163}
  {\path{doi:10.48550/ARXIV.2505.02163}}.

\bibitem{Bernevig2006}
B.~Andrei Bernevig, Taylor~L. Hughes, and Shou-Cheng Zhang.
\newblock {Quantum Spin Hall Effect and Topological Phase Transition in HgTe
  Quantum Wells}.
\newblock {\em Science}, 314(5806):1757--1761, December 2006.
\newblock \href {https://doi.org/10.1126/science.1133734}
  {\path{doi:10.1126/science.1133734}}.

\bibitem{Liu2010}
Chao-Xing Liu, Xiao-Liang Qi, HaiJun Zhang, Xi~Dai, Zhong Fang, and Shou-Cheng
  Zhang.
\newblock Model hamiltonian for topological insulators.
\newblock {\em Physical Review B}, 82(4):045122, 2010.
\newblock \href {https://doi.org/10.1103/physrevb.82.045122}
  {\path{doi:10.1103/physrevb.82.045122}}.

\bibitem{Zhao2020}
Yi-Fan Zhao, Ruoxi Zhang, Ruobing Mei, Ling-Jie Zhou, Hemian Yi, Ya-Qi Zhang,
  Jiabin Yu, Run Xiao, Ke~Wang, Nitin Samarth, Moses H.~W. Chan, Chao-Xing Liu,
  and Cui-Zu Chang.
\newblock Tuning the chern number in quantum anomalous hall insulators.
\newblock {\em Nature}, 588:419--423, 2020.
\newblock \href {https://doi.org/10.1038/s41586-020-3020-3}
  {\path{doi:10.1038/s41586-020-3020-3}}.

\bibitem{Kezilebieke2020}
Shawulienu Kezilebieke, Md~Nurul Huda, Viliam Vaňo, Markus Aapro, Somesh~C.
  Ganguli, Orlando~J. Silveira, Szczepan Głodzik, Adam~S. Foster, Teemu
  Ojanen, and Peter Liljeroth.
\newblock Topological superconductivity in a van der waals heterostructure.
\newblock {\em Nature}, 588(7838):424--428, December 2020.
\newblock \href {https://doi.org/10.1038/s41586-020-2989-y}
  {\path{doi:10.1038/s41586-020-2989-y}}.

\bibitem{Wang2021}
Yi-Xiang Wang and Fuxiang Li.
\newblock High chern number phase in topological-insulator multilayer
  structures.
\newblock {\em Physical Review B}, 104(3):035202, 2021.
\newblock \href {https://doi.org/10.1103/PhysRevB.104.035202}
  {\path{doi:10.1103/PhysRevB.104.035202}}.

\bibitem{Zhu2022}
Wenxuan Zhu, Cheng Song, Hua Bai, Liyang Liao, and Feng Pan.
\newblock High chern number quantum anomalous hall effect tunable by stacking
  order in van der waals topological insulators.
\newblock {\em Physical Review B}, 105(15):155122, April 2022.
\newblock \href {https://doi.org/10.1103/physrevb.105.155122}
  {\path{doi:10.1103/physrevb.105.155122}}.

\bibitem{Zhang2009}
Haijun Zhang, Chao-Xing Liu, Xiao-Liang Qi, Xi~Dai, Zhong Fang, and Shou-Cheng
  Zhang.
\newblock Topological insulators in bi$_2$se$_3$, bi$_2$te$_3$ and sb$_2$te$_3$
  with a single dirac cone on the surface.
\newblock {\em Nature Physics}, 5(6):438--442, 2009.
\newblock \href {https://doi.org/10.1038/nphys1270}
  {\path{doi:10.1038/nphys1270}}.

\bibitem{Yu2010}
Rui Yu, Wei Zhang, Hai-Jun Zhang, Shou-Cheng Zhang, Xi~Dai, and Zhong Fang.
\newblock Quantized anomalous hall effect in magnetic topological insulators.
\newblock {\em Science}, 329(5987):61--64, 2010.
\newblock \href {https://doi.org/10.1126/science.1187485}
  {\path{doi:10.1126/science.1187485}}.

\bibitem{Wang2013}
Jing Wang, Biao Lian, Haijun Zhang, Yong Xu, and Shou-Cheng Zhang.
\newblock Quantum anomalous hall effect with higher plateaus.
\newblock {\em Physical Review Letters}, 111(13):136801, 2013.
\newblock \href {https://doi.org/10.1103/PhysRevLett.111.136801}
  {\path{doi:10.1103/PhysRevLett.111.136801}}.

\bibitem{Shan2010}
Wen-Yu Shan, Hai-Zhou Lu, and Shun-Qing Shen.
\newblock Effective continuous model for surface states and thin films of
  three-dimensional topological insulators.
\newblock {\em New Journal of Physics}, 12(4):043048, April 2010.
\newblock \href {https://doi.org/10.1088/1367-2630/12/4/043048}
  {\path{doi:10.1088/1367-2630/12/4/043048}}.

\bibitem{Zhang2010}
Yi~Zhang, Ke~He, Cui-Zu Chang, Can-Li Song, Li-Li Wang, Xi~Chen, Jin-Feng Jia,
  Zhong Fang, Xi~Dai, Wen-Yu Shan, Shun-Qing Shen, Qian Niu, Xiao-Liang Qi,
  Shou-Cheng Zhang, Xu-Cun Ma, and Qi-Kun Xue.
\newblock Crossover of the three-dimensional topological insulator bi2se3 to
  the two-dimensional limit.
\newblock {\em Nature Physics}, 6(8):584--588, 2010.
\newblock \href {https://doi.org/10.1038/nphys1689}
  {\path{doi:10.1038/nphys1689}}.

\bibitem{Zhang2014}
Shu-feng Zhang, Hua Jiang, X.~C. Xie, and Qing-feng Sun.
\newblock Effect of magnetic field on a magnetic topological insulator film
  with structural inversion asymmetry.
\newblock {\em Physical Review B}, 89(15):155419, April 2014.
\newblock \href {https://doi.org/10.1103/physrevb.89.155419}
  {\path{doi:10.1103/physrevb.89.155419}}.

\bibitem{Schnyder2008}
Andreas~P Schnyder, Shinsei Ryu, Akira Furusaki, and Andreas W.~W. Ludwig.
\newblock Classification of topological insulators and superconductors in three
  spatial dimensions.
\newblock {\em Physical Review B}, 78(19):195125, 2008.
\newblock \href {https://doi.org/10.1103/PhysRevB.78.195125}
  {\path{doi:10.1103/PhysRevB.78.195125}}.

\bibitem{Chiu2016}
Ching-Kai Chiu, Jeffrey C-Y Teo, Andreas~P Schnyder, and Shinsei Ryu.
\newblock Classification of topological quantum matter with symmetries.
\newblock {\em Reviews of Modern Physics}, 88(3):035005, 2016.
\newblock \href {https://doi.org/10.1103/RevModPhys.88.035005}
  {\path{doi:10.1103/RevModPhys.88.035005}}.

\bibitem{Sticlet2012}
Doru Sticlet, Cristina Bena, and Pascal Simon.
\newblock Spin and majorana polarization in topological superconducting wires.
\newblock {\em Physical Review Letters}, 108(9):096802, March 2012.
\newblock \href {https://doi.org/10.1103/physrevlett.108.096802}
  {\path{doi:10.1103/physrevlett.108.096802}}.

\bibitem{Sedlmayr2015}
N.~Sedlmayr and C.~Bena.
\newblock Visualizing majorana bound states in one and two dimensions using the
  generalized majorana polarization.
\newblock {\em Phys. Rev. B}, 92:115115, Sep 2015.
\newblock \href {https://doi.org/10.1103/PhysRevB.92.115115}
  {\path{doi:10.1103/PhysRevB.92.115115}}.

\bibitem{Karoliya2025}
Shubhanshu Karoliya, Sumanta Tewari, and Gargee Sharma.
\newblock Majorana polarization in disordered quasi-one-dimensional hybrid
  nanowires.
\newblock {\em Phys. Rev. B}, 112:165410, Oct 2025.
\newblock \href {https://doi.org/10.1103/7lh9-7gsq}
  {\path{doi:10.1103/7lh9-7gsq}}.

\bibitem{Novik2020}
E.~G. Novik, B.~Trauzettel, and P.~Recher.
\newblock Transport signatures of a junction between a quantum spin hall system
  and a chiral topological superconductor.
\newblock {\em Physical Review B}, 101(23):235308, 2020.
\newblock \href {https://doi.org/10.1103/PhysRevB.101.235308}
  {\path{doi:10.1103/PhysRevB.101.235308}}.

\bibitem{Wang2014}
Jing Wang, Yong Xu, and Shou-Cheng Zhang.
\newblock Two-dimensional time-reversal-invariant topological superconductivity
  in a doped quantum spin-hall insulator.
\newblock {\em Physical Review B}, 90(5):054503, August 2014.
\newblock \href {https://doi.org/10.1103/physrevb.90.054503}
  {\path{doi:10.1103/physrevb.90.054503}}.

\bibitem{Ji2024}
Haijiao Ji and Noah F.~Q. Yuan.
\newblock Superconducting properties of bernevig-hughes-zhang model: Theory and
  applications to transition metal dichalcogenides.
\newblock {\em Physical Review B}, 109(5):054510, February 2024.
\newblock \href {https://doi.org/10.1103/physrevb.109.054510}
  {\path{doi:10.1103/physrevb.109.054510}}.

\bibitem{Baba2022}
Yuriko Baba, Mario Amado, Enrique Diez, Francisco Domínguez-Adame, and
  Rafael~A. Molina.
\newblock Effect of external fields in high-chern-number quantum anomalous hall
  insulators.
\newblock {\em Physical Review B}, 106(24):245305, 2022.
\newblock \href {https://doi.org/10.1103/PhysRevB.106.245305}
  {\path{doi:10.1103/PhysRevB.106.245305}}.

\bibitem{Zhao2023}
Yi‐Fan Zhao, Ruoxi Zhang, Zi‐Ting Sun, Ling‐Jie Zhou, Deyi Zhuo, Zi‐Jie
  Yan, Hemian Yi, Ke~Wang, Moses H.~W. Chan, Chao‐Xing Liu, K.~T. Law, and
  Cui‐Zu Chang.
\newblock 3d quantum anomalous hall effect in magnetic topological insulator
  trilayers of hundred‐nanometer thickness.
\newblock {\em Advanced Materials}, 36(13), December 2023.
\newblock \href {https://doi.org/10.1002/adma.202310249}
  {\path{doi:10.1002/adma.202310249}}.

\bibitem{Ovchinnikov2021}
Dmitry Ovchinnikov, Xiong Huang, Zhong Lin, Zaiyao Fei, Jiaqi Cai, Tiancheng
  Song, Minhao He, Qianni Jiang, Chong Wang, Hao Li, Yayu Wang, Yang Wu,
  Di~Xiao, Jiun-Haw Chu, Jiaqiang Yan, Cui-Zu Chang, Yong-Tao Cui, and Xiaodong
  Xu.
\newblock Intertwined topological and magnetic orders in atomically thin chern
  insulator mnbi\textsubscript{2}te\textsubscript{4}.
\newblock {\em Nano Letters}, 21(6):2544--2550, March 2021.
\newblock \href {https://doi.org/10.1021/acs.nanolett.0c05117}
  {\path{doi:10.1021/acs.nanolett.0c05117}}.

\bibitem{Mei2024}
Ruobing Mei, Yi-Fan Zhao, Chong Wang, Yafei Ren, Di~Xiao, Cui-Zu Chang, and
  Chao-Xing Liu.
\newblock Electrically controlled anomalous hall effect and orbital
  magnetization in topological magnet mnbi2te4.
\newblock {\em Physical Review Letters}, 132(6):066604, February 2024.
\newblock \href {https://doi.org/10.1103/physrevlett.132.066604}
  {\path{doi:10.1103/physrevlett.132.066604}}.

\bibitem{Yuan2024}
Wei Yuan, Zi-Jie Yan, Hemian Yi, Zihao Wang, Stephen Paolini, Yi-Fan Zhao,
  Lingjie Zhou, Annie~G. Wang, Ke~Wang, Thomas Prokscha, Zaher Salman, Andreas
  Suter, Purnima~P. Balakrishnan, Alexander~J. Grutter, Laurel~E. Winter, John
  Singleton, Moses H.~W. Chan, and Cui-Zu Chang.
\newblock Coexistence of superconductivity and antiferromagnetism in
  topological magnet mnbi\textsubscript{2}te\textsubscript{4} films.
\newblock {\em Nano Letters}, 24(26):7962--7971, 2024.
\newblock \href {https://doi.org/10.1021/acs.nanolett.4c01407}
  {\path{doi:10.1021/acs.nanolett.4c01407}}.

\bibitem{Vyazovskaya2025}
Alexandra~Yu. Vyazovskaya, Mihovil Bosnar, Evgueni~V. Chulkov, and Mikhail~M.
  Otrokov.
\newblock Intrinsic magnetic topological insulators of the mnbi2te4 family.
\newblock {\em Communications Materials}, 6(1), April 2025.
\newblock \href {https://doi.org/10.1038/s43246-025-00794-3}
  {\path{doi:10.1038/s43246-025-00794-3}}.

\end{thebibliography}
\end{document}